\documentclass[11pt]{article}
\usepackage{graphicx}
\usepackage{epsfig}
\usepackage{epsf}
\usepackage{epstopdf}

\begin{document}
\pagestyle{empty}
\def\eqa{\!\!&=&\!\!}
\def\ccr{\nonumber\\}

\def\la{\langle}
\def\ra{\rangle}

\def\del{\Delta}
\def\ddel{{}^\bullet\! \Delta}
\def\deld{\Delta^{\hskip -.5mm \bullet}}
\def\ddeld{{}^{\bullet}\! \Delta^{\hskip -.5mm \bullet}}
\def\dddel{{}^{\bullet \bullet} \! \Delta}

\def\rld{\rlap{\,/}D}
\def\rldd{\rlap{\,/}\nabla}
%
\def\half{{1\over 2}}
\def\third{{1\over3}}
\def\fourth{{1\over4}}
\def\fifth{{1\over5}}
\def\sixth{{1\over6}}
\def\seventh{{1\over7}}
\def\eigth{{1\over8}}
\def\ninth{{1\over9}}
\def\tenth{{1\over10}}
\def\bN{\mathop{\bf N}}
\def\R{{\rm I\!R}}
\def\Eins{{\mathchoice {\rm 1\mskip-4mu l} {\rm 1\mskip-4mu l}
{\rm 1\mskip-4.5mu l} {\rm 1\mskip-5mu l}}}
\def\Z{{\mathchoice {\hbox{$\sf\textstyle Z\kern-0.4em Z$}}
{\hbox{$\sf\textstyle Z\kern-0.4em Z$}}
{\hbox{$\sf\scriptstyle Z\kern-0.3em Z$}}
{\hbox{$\sf\scriptscriptstyle Z\kern-0.2em Z$}}}}
\def\abs#1{\left| #1\right|}
\def\com#1#2{
        \left[#1, #2\right]}
\def\square{\kern1pt\vbox{\hrule height 1.2pt\hbox{\vrule width 1.2pt
   \hskip 3pt\vbox{\vskip 6pt}\hskip 3pt\vrule width 0.6pt}
   \hrule height 0.6pt}\kern1pt}
      \def\boxop{{\raise-.25ex\hbox{\square}}}
\def\contract{\makebox[1.2em][c]{
        \mbox{\rule{.6em}{.01truein}\rule{.01truein}{.6em}}}}
\def\ltap{\ \raisebox{-.4ex}{\rlap{$\sim$}} \raisebox{.4ex}{$<$}\ }
\def\gtap{\ \raisebox{-.4ex}{\rlap{$\sim$}} \raisebox{.4ex}{$>$}\ }
\def\mn{{\mu\nu}}
\def\rs{{\rho\sigma}}
\newcommand{\Det}{{\rm Det}}
\def\Tr{{\rm Tr}\,}
\def\tr{{\rm tr}\,}
\def\sumij{\sum_{i<j}}
\def\e{\,{\rm e}}
\def\pa{\partial}
\def\dA{\partial^2}
\def\ddx{{d\over dx}}
\def\ddt{{d\over dt}}
\def\der#1#2{{d #1\over d#2}}
\def\lie{\hbox{\it \$}} 
\def\partder#1#2{{\partial #1\over\partial #2}}
\def\secder#1#2#3{{\partial^2 #1\over\partial #2 \partial #3}}
%
\newcommand{\be}{\begin{equation}}
\newcommand{\ee}{\end{equation}\noindent}
\newcommand{\bear}{\begin{eqnarray}}
\newcommand{\ear}{\end{eqnarray}\noindent}
\newcommand{\benn}{\begin{enumerate}}
\newcommand{\enn}{\end{enumerate}}
\newcommand{\veject}{\vfill\eject}
\newcommand{\ven}{\vfill\eject\noindent}
%
\def\eq#1{{eq. (\ref{#1})}}
\def\eqs#1#2{{eqs. (\ref{#1}) -- (\ref{#2})}}
%
\def\totint{\int_{-\infty}^{\infty}}
\def\posint{\int_0^{\infty}}
\def\negint{\int_{-\infty}^0}
\def\pint{{\dps\int}{dp_i\over {(2\pi)}^d}}
%
\newcommand{\GeV}{\mbox{GeV}}
\def\FFdual{F\cdot\tilde F}
\def\bra#1{\langle #1 |}
\def\ket#1{| #1 \rangle}
\def\braket#1#2{\langle {#1} \mid {#2} \rangle}
\def\vev#1{\langle #1 \rangle}
\def\rightvac{\mid 0\rangle}
\def\leftvac{\langle 0\mid}
\def\ihbar{{i\over\hbar}}
\def\ge{\hbox{$\gamma_1$}}
\def\gz{\hbox{$\gamma_2$}}
\def\gd{\hbox{$\gamma_3$}}
\def\go{\hbox{$\gamma_1$}}
\def\gt{\hbox{\$\gamma_2$}}
\def\gth{\hbox{$\gamma_3$}} 
\def\gf{\hbox{$\gamma_5\;$}}
\def\slash#1{#1\!\!\!\raise.15ex\hbox {/}}
\newcommand{\slD}{\,\raise.15ex\hbox{$/$}\kern-.27em\hbox{$\!\!\!D$}}
\newcommand{\slpartial}{\raise.15ex\hbox{$/$}\kern-.57em\hbox{$\partial$}}
\newcommand{\cL}{\cal L}
\newcommand{\D}{\cal D}
\newcommand{\Dhalf}{{D\over 2}}
\def\eps{\epsilon}
\def\epshalf{{\epsilon\over 2}}
\def\lag{( -\partial^2 + V)}
\def\freeexp{{\rm e}^{-\int_0^Td\tau {1\over 4}\dot x^2}}
\def\kinb{{1\over 4}\dot x^2}
\def\kinf{{1\over 2}\psi\dot\psi}
\def\expk{{\rm exp}\biggl[\,\sum_{i<j=1}^4 G_{Bij}k_i\cdot k_j\biggr]}
\def\expp{{\rm exp}\biggl[\,\sum_{i<j=1}^4 G_{Bij}p_i\cdot p_j\biggr]}
\def\expshort{{\e}^{\half G_{Bij}k_i\cdot k_j}}
\def\expabb{{\e}^{(\cdot )}}
\def\epseps#1#2{\varepsilon_{#1}\cdot \varepsilon_{#2}}
\def\epsk#1#2{\varepsilon_{#1}\cdot k_{#2}}
\def\kk#1#2{k_{#1}\cdot k_{#2}}
\def\G#1#2{G_{B#1#2}}
\def\Gp#1#2{{\dot G_{B#1#2}}}
\def\GF#1#2{G_{F#1#2}}
\def\Dab{{(x_a-x_b)}}
\def\Dsq{{({(x_a-x_b)}^2)}}
\def\PITD{{(4\pi T)}^{-{D\over 2}}}
\def\4piTD{{(4\pi T)}^{-{D\over 2}}}
\def\4piT4{{(4\pi T)}^{-2}}
\def\TintmD{{\dps\int_{0}^{\infty}}{dT\over T}\,e^{-m^2T}
    {(4\pi T)}^{-{D\over 2}}}
\def\Tintm4{{\dps\int_{0}^{\infty}}{dT\over T}\,e^{-m^2T}
    {(4\pi T)}^{-2}}
\def\Tintm{{\dps\int_{0}^{\infty}}{dT\over T}\,e^{-m^2T}}
\def\Tint{{\dps\int_{0}^{\infty}}{dT\over T}}
\def\np{n_{+}}
\def\nm{n_{-}}
\def\Np{N_{+}}
\def\Nm{N_{-}}
\newcommand{\slG}{{{\dot G}\!\!\!\! \raise.15ex\hbox {/}}}
\newcommand{\Gd}{{\dot G}}
\newcommand{\Gund}{{\underline{\dot G}}}
\newcommand{\Gdd}{{\ddot G}}
\def\GBd12{{\dot G}_{B12}}
\def\Dx{\dps\int{\cal D}x}
\def\Dy{\dps\int{\cal D}y}
\def\Dpsi{\dps\int{\cal D}\psi}
\def\dint#1{\int\!\!\!\!\!\int\limits_{\!\!#1}}
\def\ddtau{{d\over d\tau}}
\def\ie{\hbox{$\textstyle{\int_1}$}}
\def\iz{\hbox{$\textstyle{\int_2}$}}
\def\id{\hbox{$\textstyle{\int_3}$}}
\def\ldop{\hbox{$\lbrace\mskip -4.5mu\mid$}}
\def\rdop{\hbox{$\mid\mskip -4.3mu\rbrace$}}
%
\newcommand{\1}{{\'\i}}
\newcommand{\no}{\noindent}
\def\non{\nonumber}
\def\dps{\displaystyle}
\def\sy{\scriptscriptstyle}
\def\sy{\scriptscriptstyle}

%

\newcommand{\bea}{\begin{eqnarray}}  
\newcommand{\eea}{\end{eqnarray}}  
\def\eqa{&=&}  
\def\ccr{\nonumber\\}  
  
\def\a{\alpha}
\def\b{\beta}
\def\m{\mu}
\def\n{\nu}
\def\r{\rho}
\def\s{\sigma}
\def\ep{\epsilon}

\def\cosech{\rm cosech}
\def\sech{\rm sech}
\def\coth{\rm coth}
\def\tanh{\rm tanh}

\def\sqr#1#2{{\vcenter{\vbox{\hrule height.#2pt  
     \hbox{\vrule width.#2pt height#1pt \kern#1pt  
           \vrule width.#2pt}  
       \hrule height.#2pt}}}}  
\def\square{\mathchoice\sqr66\sqr66\sqr{2.1}3\sqr{1.5}3}  
  
\def\appendix{\par\clearpage
  \setcounter{section}{0}
  \setcounter{subsection}{0}
  \def\@sectname{Appendix~}
  \def\theequation{\thesection\arabic{equation}}
  \def\thesection{\Alph{section}}}
 
\def\thefigures#1{\par\clearpage\section*{Figures\@mkboth
  {FIGURES}{FIGURES}}\list
  {Fig.~\arabic{enumi}.}{\labelwidth\parindent\advance
\labelwidth -\labelsep
      \leftmargin\parindent\usecounter{enumi}}}
\def\figitem#1{\item\label{#1}}
\let\endthefigures=\endlist
 
\def\thetables#1{\par\clearpage\section*{Tables\@mkboth
  {TABLES}{TABLES}}\list
  {Table~\Roman{enumi}.}{\labelwidth-\labelsep
      \leftmargin0pt\usecounter{enumi}}}
\def\tableitem#1{\item\label{#1}}
\let\endthetables=\endlist
 
\def\@sect#1#2#3#4#5#6[#7]#8{\ifnum #2>\c@secnumdepth
     \def\@svsec{}\else
     \refstepcounter{#1}\edef\@svsec{\@sectname\csname the#1\endcsname
.\hskip 1em }\fi
     \@tempskipa #5\relax
      \ifdim \@tempskipa>\z@
        \begingroup #6\relax
          \@hangfrom{\hskip #3\relax\@svsec}{\interlinepenalty \@M #8\par}
        \endgroup
       \csname #1mark\endcsname{#7}\addcontentsline
         {toc}{#1}{\ifnum #2>\c@secnumdepth \else
                      \protect\numberline{\csname the#1\endcsname}\fi
                    #7}\else
        \def\@svse=chd{#6\hskip #3\@svsec #8\csname #1mark\endcsname
                      {#7}\addcontentsline
                           {toc}{#1}{\ifnum #2>\c@secnumdepth \else
                             \protect\numberline{\csname the#1\endcsname}\fi
                       #7}}\fi
     \@xsect{#5}}
 
\def\@sectname{}
%
%
\def\eg{\hbox{\it e.g.}}        \def\cf{\hbox{\it cf.}}
\def\etal{\hbox{\it et al.}}
\def\dash{\hbox{---}}
\def\bR{\mathop{\bf R}}
\def\bC{\mathop{\bf C}}
\def\eq#1{{eq. \ref{#1}}}
\def\eqs#1#2{{eqs. \ref{#1}--\ref{#2}}}
\def\lie{\hbox{\it \$}} 
\def\partder#1#2{{\partial #1\over\partial #2}}
\def\secder#1#2#3{{\partial^2 #1\over\partial #2 \partial #3}}
\def\abs#1{\left| #1\right|}
\def\ltap{\ \raisebox{-.4ex}{\rlap{$\sim$}} \raisebox{.4ex}{$<$}\ }
\def\gtap{\ \raisebox{-.4ex}{\rlap{$\sim$}} \raisebox{.4ex}{$>$}\ }
\def\contract{\makebox[1.2em][c]{
        \mbox{\rule{.6em}{.01truein}\rule{.01truein}{.6em}}}}
%
\def\com#1#2{
        \left[#1, #2\right]}
%
%
\def\bentarrow{\:\raisebox{1.3ex}{\rlap{$\vert$}}\!\rightarrow}
\def\longbent{\:\raisebox{3.5ex}{\rlap{$\vert$}}\raisebox{1.3ex}%
        {\rlap{$\vert$}}\!\rightarrow}
\def\onedk#1#2{
        \begin{equation}
        \begin{array}{l}
         #1 \\
         \bentarrow #2
        \end{array}
        \end{equation}
                }
\def\dk#1#2#3{
        \begin{equation}
        \begin{array}{r c l}
        #1 & \rightarrow & #2 \\
         & & \bentarrow #3
        \end{array}
        \end{equation}
                }
\def\dkp#1#2#3#4{
        \begin{equation}
        \begin{array}{r c l}
        #1 & \rightarrow & #2#3 \\
         & & \phantom{\; #2}\bentarrow #4
        \end{array}
        \end{equation}
                }
\def\bothdk#1#2#3#4#5{
        \begin{equation}
        \begin{array}{r c l}
        #1 & \rightarrow & #2#3 \\
         & & \:\raisebox{1.3ex}{\rlap{$\vert$}}\raisebox{-0.5ex}{$\vert$}%
        \phantom{#2}\!\bentarrow #4 \\
         & & \bentarrow #5
        \end{array}
        \end{equation}
                }
\newcommand{\nc}{\newcommand}
\nc{\spa}[3]{\left\langle#1\,#3\right\rangle}
\nc{\spb}[3]{\left[#1\,#3\right]}
\nc{\ksl}{\not{\hbox{\kern-2.3pt $k$}}}
\nc{\hf}{\textstyle{1\over2}}
\nc{\pol}{\varepsilon}
\nc{\tq}{{\tilde q}}
\nc{\esl}{\not{\hbox{\kern-2.3pt $\pol$}}}
\renewcommand{\theequation}{\arabic{section}.\arabic{equation}}
\renewcommand{\arraystretch}{2.5}
\def\R{1\!\!{\rm R}}
\def\Eins{\mathord{1\hskip -1.5pt
\vrule width .5pt height 7.75pt depth -.2pt \hskip -1.2pt
\vrule width 2.5pt height .3pt depth -.05pt \hskip 1.5pt}}
\newcommand{\symb}{\mbox{symb}}
\renewcommand{\arraystretch}{2.5}
\def\GBd12{{\dot G}_{B12}}
\def\mneg{\!\!\!\!\!\!\!\!\!\!}
\def\Mneg{\!\!\!\!\!\!\!\!\!\!\!\!\!\!\!\!\!\!\!\!}
\def\non{\nonumber}
\def\beqn*{\begin{eqnarray*}}
\def\eqn*{\end{eqnarray*}}
\def\sy{\scriptscriptstyle}
\def\footstrut{\baselineskip 12pt}
\def\square{\kern1pt\vbox{\hrule height 1.2pt\hbox{\vrule width 1.2pt
   \hskip 3pt\vbox{\vskip 6pt}\hskip 3pt\vrule width 0.6pt}
   \hrule height 0.6pt}\kern1pt}
\def\np{n_{+}}
\def\nm{n_{-}}
\def\Np{N_{+}}
\def\Nm{N_{-}}
\def\exmn{\Bigl(\mu \leftrightarrow \nu \Bigr)}
\def\slash#1{#1\!\!\!\raise.15ex\hbox {/}}
\def\dint#1{\int\!\!\!\!\!\int\limits_{\!\!#1}}
\def\bra#1{\langle #1 |}
\def\ket#1{| #1 \rangle}
\def\vev#1{\langle #1 \rangle}
\def\rightvac{\mid 0\rangle}
\def\leftvac{\langle 0\mid}
\def\dps{\displaystyle}
\def\sy{\scriptscriptstyle}
\def\half{{1\over 2}}
\def\third{{1\over3}}
\def\fourth{{1\over4}}
\def\fifth{{1\over5}}
\def\sixth{{1\over6}}
\def\seventh{{1\over7}}
\def\eigth{{1\over8}}
\def\ninth{{1\over9}}
\def\tenth{{1\over10}}
\def\pa{\partial}
\def\ddtau{{d\over d\tau}}
\def\ge{\hbox{\textfont1=\tame $\gamma_1$}}
\def\gz{\hbox{\textfont1=\tame $\gamma_2$}}
\def\gd{\hbox{\textfont1=\tame $\gamma_3$}}
\def\go{\hbox{\textfont1=\tamt $\gamma_1$}}
\def\gt{\hbox{\textfont1=\tamt $\gamma_2$}}
\def\gth{\hbox{\textfont1=\tamt $\gamma_3$}} 
\def\gf{\hbox{$\gamma_5\;$}}
\def\ie{\hbox{$\textstyle{\int_1}$}}
\def\iz{\hbox{$\textstyle{\int_2}$}}
\def\id{\hbox{$\textstyle{\int_3}$}}
\def\ldop{\hbox{$\lbrace\mskip -4.5mu\mid$}}
\def\rdop{\hbox{$\mid\mskip -4.3mu\rbrace$}}
\def\eps{\epsilon}
\def\epshalf{{\epsilon\over 2}}
\def\e{\mbox{e}}
\def\mn{{\mu\nu}}
\def\exmn{{(\mu\leftrightarrow\nu )}}
\def\ab{{\alpha\beta}}
\def\exab{{(\alpha\leftrightarrow\beta )}}
\def\g{\mbox{g}}
\def\kinb{{1\over 4}\dot x^2}
\def\kinf{{1\over 2}\psi\dot\psi}
\def\expk{{\rm exp}\biggl[\,\sum_{i<j=1}^4 G_{Bij}k_i\cdot k_j\biggr]}
\def\expp{{\rm exp}\biggl[\,\sum_{i<j=1}^4 G_{Bij}p_i\cdot p_j\biggr]}
\def\expshort{{\e}^{\half G_{Bij}k_i\cdot k_j}}
\def\expabb{{\e}^{(\cdot )}}
\def\epseps#1#2{\varepsilon_{#1}\cdot \varepsilon_{#2}}
\def\epsk#1#2{\varepsilon_{#1}\cdot k_{#2}}
\def\kk#1#2{k_{#1}\cdot k_{#2}}
\def\G#1#2{G_{B#1#2}}
\def\Gp#1#2{{\dot G_{B#1#2}}}
\def\GF#1#2{G_{F#1#2}}
\def\Dab{{(x_a-x_b)}}
\def\Dsq{{({(x_a-x_b)}^2)}}
\def\lag{( -\partial^2 + V)}
\def\PITD{{(4\pi T)}^{-{D\over 2}}}
\def\4piTD{{(4\pi T)}^{-{D\over 2}}}
\def\4piT4{{(4\pi T)}^{-2}}
\def\TintmD{{\dps\int_{0}^{\infty}}{dT\over T}\,e^{-m^2T}
    {(4\pi T)}^{-{D\over 2}}}
\def\Tintm4{{\dps\int_{0}^{\infty}}{dT\over T}\,e^{-m^2T}
    {(4\pi T)}^{-2}}
\def\Tintm{{\dps\int_{0}^{\infty}}{dT\over T}\,e^{-m^2T}}
\def\Tint{{\dps\int_{0}^{\infty}}{dT\over T}}
\def\pint{{\dps\int}{dp_i\over {(2\pi)}^d}}
\def\Dx{\dps\int{\cal D}x}
\def\Dy{\dps\int{\cal D}y}
\def\Dpsi{\dps\int{\cal D}\psi}
\def\Tr{{\rm Tr}\,}
\def\tr{{\rm tr}\,}
\def\sumij{\sum_{i<j}}
\def\freeexp{{\rm e}^{-\int_0^Td\tau {1\over 4}\dot x^2}}
\def\arraystretch{2.5}
\def\Ge{\mbox{GeV}}
\def\dA{\partial^2}
\def\DA{\sqsubset\!\!\!\!\sqsupset}
\def\FFdual{F\cdot\tilde F}
\def\mn{{\mu\nu}}
\def\rs{{\rho\sigma}}
\def\oplusotimes{{{\lower 15pt\hbox{$\scriptscriptstyle \oplus$}}\atop{\otimes}}}
\def\perppar{{{\lower 15pt\hbox{$\scriptscriptstyle \perp$}}\atop{\parallel}}}
\def\oopp{{{\lower 15pt\hbox{$\scriptscriptstyle \oplus$}}\atop{\otimes}}\!{{\lower 15pt\hbox{$\scriptscriptstyle \perp$}}\atop{\parallel}}}
%
%
\def\bbbr{{\rm I\!R}}
\def\bbbone{{\mathchoice {\rm 1\mskip-4mu l} {\rm 1\mskip-4mu l}
{\rm 1\mskip-4.5mu l} {\rm 1\mskip-5mu l}}}
\def\bbbz{{\mathchoice {\hbox{$\sf\textstyle Z\kern-0.4em Z$}}
{\hbox{$\sf\textstyle Z\kern-0.4em Z$}}
{\hbox{$\sf\scriptstyle Z\kern-0.3em Z$}}
{\hbox{$\sf\scriptscriptstyle Z\kern-0.2em Z$}}}}

\renewcommand{\thefootnote}{\protect\arabic{footnote}}
\hfill {\large AEI-2008-053}
\vskip 0.0cm
\hfill {\large UMSNH-IFM-F-2008-24}

\begin{center}
{\huge\bf Gravitational corrections to the}
\vspace{5pt}

{\huge\bf Euler-Heisenberg Lagrangian}
\vskip1.3cm

{\large Fiorenzo Bastianelli$^{a,b}$, Jos\'e Manuel D\'avila$^{c}$, 
Christian Schubert$^{a,c}$}
\\[1.5ex]

\begin{itemize}
\item [$^a$]
{\it 
Max-Planck-Institut f\"ur Gravitationsphysik, Albert-Einstein-Institut,
M\"uhlenberg 1, D-14476 Potsdam, Germany
}
\item [$^b$]
{\it
Dipartimento di Fisica, Universit\`a di Bologna and INFN, Sezione di Bologna,
Via Irnerio 46, I-40126 Bologna, Italy
}
\item [$^c$]
{\it 
Instituto de F\'{\i}sica y Matem\'aticas
\\
Universidad Michoacana de San Nicol\'as de Hidalgo\\
Edificio C-3, Apdo. Postal 2-82\\
C.P. 58040, Morelia, Michoac\'an, M\'exico\\
}
\end{itemize}
\end{center}
\centerline{\today}
\vspace{1cm}
 {\large \bf Abstract:}
We use the worldline formalism for calculating the one-loop effective action
for the Einstein-Maxwell background induced by charged scalars or spinors, 
in the limit of low energy and weak
gravitational field but treating the electromagnetic field nonperturbatively.
The effective action is obtained in a form which generalizes the standard
proper-time representation of the Euler-Heisenberg Lagrangian. 
We compare with previous work and discuss possible applications. 
\begin{quotation}

\end{quotation}
\vfill\eject
\pagestyle{plain}
\setcounter{page}{1}
\setcounter{footnote}{0}

\vspace{10pt}
\section{Introduction}
\renewcommand{\theequation}{1.\arabic{equation}}
\setcounter{equation}{0}

In 1936 Heisenberg and Euler derived their
famous effective Lagrangian \cite{eulhei}
describing the effect of a virtual electron - positron pair
on an external Maxwell field in the one loop and constant field
approximation. Its standard proper time representation is

\bear
{\cal L}_{\rm spin}(F)&=& - {1\over 8\pi^2}
\int_0^{\infty}{dT\over T^3}
\,\e^{-m^2T}
\biggl\lbrack
{(eaT)(ebT)\over {\rm tanh}(eaT)\tan(ebT)} 
\nonumber\\&&\hspace{70pt}
- {e^2\over 3}(a^2-b^2)T^2 -1
\biggr\rbrack \, .
\label{ehspin}
\ear
Here $T$ is the proper-time of the loop fermion, $m$ its mass, and $a,b$ are the two
invariants of the Maxwell field, 
related to $\bf E$, $\bf B$ by $a^2-b^2 = B^2-E^2,\quad ab = {\bf E}\cdot {\bf B}$.
The two subtraction terms implement the renormalization of charge and vacuum
energy.

An analogous representation was found later for scalar QED
\cite{weisskopf,schwinger51}:

\bear
{\cal L}_{\rm scal}(F)&=&  {1\over 16\pi^2}
\int_0^{\infty}{dT\over T^3}
\,\e^{-m^2T}
\biggl[
{(eaT)(ebT)\over \sinh(eaT)\sin(ebT)} 
\nonumber\\&&\hspace{60pt}
+{e^2\over 6}(a^2-b^2)T^2 -1
\biggr] \, .
\label{ehscal}
\ear
Although the effective Lagrangian for scalar QED is due to Weisskopf
and Schwinger, for simplicity we will call it the ``Scalar Euler-Heisenberg Lagrangian''.
 
The Lagrangians (\ref{ehspin}), (\ref{ehscal})
historically provided the first examples for the concept of an
effective Lagrangian, as well as the first nonperturbative
results in quantum field theory. 
Despite of their formal simplicity they contain an enormous amount of
physical information on low energy processes in QED. 
See \cite{ditreu,ditgiebook,geraldrev} for reviews of their various applications
and generalizations.

The proper time integrals in eqs. (\ref{ehspin}), (\ref{ehscal}) 
can be done exactly
in terms of certain special functions \cite{geraldrev}. Alternatively,
one can expand the integrands as power series in the field invariants,
using the Taylor expansions
\bear
{z \over\ {\rm tanh}(z)} &=& \sum_{n=0}^{\infty}{B_{2n}\over
(2n)!} \,(2z)^{2n} \, , \label{taylcoth2}\\
{z\over {\rm sinh}(z)} &=&
-\sum_{n=0}^{\infty}\Bigl(1-2^{1-2n}\Bigr){B_{2n}\over (2n)!}\,(2z)^{2n} \, .
\label{taylcsch}
\ear
Here the $B_{2n}$ are Bernoulli numbers.
The terms in this expansion involving $N=2n$ powers of the field
contain the information on the low energy limit of the $N$ photon scattering
amplitudes, defined by all photon energies being small compared to the loop mass,
$\omega_i \ll m, i=1,\ldots, N$.   
Thus in this limit the effective Lagrangian
allows one to obtain these amplitudes in closed form \cite{mascvi}. 
This should be
contrasted with the fact that, away from this limit, the
$N$ photon amplitudes are still poorly known. 
The four photon scattering amplitudes were obtained a long time
ago \cite{fourphoton}, but the explicit calculation for the six-point case
became possible only recently \cite{sixphoton}. Beyond six points, only the
maximally helicity violating $N$ - photon amplitude ($N-2$ equal helicities) 
has been obtained so far \cite{mahlon,babjva}. 
Regarding the off-shell case, to the best of our knowledge even the
four-point amplitude is known only with maximally two
legs off-shell \cite{cotopi}. 

Apart from the purely magnetic field case, the Euler-Heisenberg Lagrangians 
have also imaginary parts, induced by the poles which the integrands in 
(\ref{ehspin}), (\ref{ehscal}) have for $b\ne 0$. A simple application of 
Cauchy's theorem yields Schwinger's representation \cite{schwinger51}

\bear
{\rm Im} {\cal L}_{\rm spin}(E) &=&  \frac{m^4}{8\pi^3}
\beta^2\, \sum_{k=1}^\infty \frac{1}{k^2}
\,\exp\left[-\frac{\pi k}{\beta}\right]  \, , \non\\
{\rm Im} {\cal L}_{\rm scal}(E) &=&  \frac{m^4}{16\pi^3}
\beta^2\, \sum_{k=1}^\infty \frac{(-1)^{k+1}}{k^2} 
\,\exp\left[-\frac{\pi k}{\beta}\right]  \non\\
\label{ImL}
\ear
with $\beta = {eE\over m^2}$. These imaginary parts directly relate to
the rates of electron-positron pair production by the electric field \cite{schwinger51,geraldrev}.
The representation (\ref{ImL}) makes it clear that this effect is nonperturbative in the
field; its calculation requires the knowledge of the effective action to all orders
in the weak field expansion. 

Concerning higher loop corrections to the Lagrangians (\ref{ehspin}),
(\ref{ehscal}) see
\cite{ritspin,ditreu, rescsc,frss} for the spinor and \cite{ritscal,rescsc,frss,afalma} 
for the scalar case.

In the present article, we will study the corrections to the Euler-Heisenberg Lagrangians
(\ref{ehspin}), (\ref{ehscal}) due to an additional weak gravitational background. 
This amounts to calculating the one-loop
effective actions for a generic Einstein-Maxwell background due to a scalar or spinor loop,
to all orders in the electromagnetic field strength, and to leading order in the curvature.

A sizable body of work exists already on the one-loop effective action
in mixed gravitational-electromagnetic fields.  
Drummond and Hathrell in their seminal work \cite{druhat} obtained 
the terms in the fermionic effective Lagrangian involving one curvature tensor and
two field strength tensors:

\bear
{\cal L}_{\rm spin}^{(DH)} &=& 
\frac{1}{180 (4\pi)^2m^2} \bigg(
5 R F_{\mu\nu}^2 
-26 R_{\mu\nu} F^{\mu\alpha} F^\nu{}_\alpha
+2  R_{\mu\nu\alpha\beta}F^{\mu\nu}F^{\alpha\beta} \non\\
&& \qquad\qquad
+24 (\nabla^\alpha F_{\alpha\mu})^2  
\bigg )
\label{drumhath}
\ear
(here and in the following we will often absorb the electric charge $e$ into the field strength tensor
$F$).
The motivation of \cite{druhat} for considering these terms was that they contain information
on the modification of the photon dispersion relation by a generic gravitational
background. While it is well-known that such modifications exist already in the
pure QED case \cite{ditgiebook}, 
the gravitational case is particularly interesting in that it
permits superluminal propagation \cite{lapata,ditgie}, leading even to 
speculations on a possible violation of microcausality \cite{shore}. 
However, as emphasized in \cite{holsho} these issues cannot be resolved at the
level of the low-energy effective action since this would require information on the photon
propagation in the full energy range. 

As usual, a systematic computation of this effective action for either the scalar or spinor loop
cases requires one to decide on the grouping of terms, the three basic options being

\benn

\item
Summing over all derivatives on fields with the number of fields fixed.

\item
Grouping together terms with a fixed mass dimension.

\item
Fixing the number of derivatives and summing over the number of fields.

\enn
The first approach is usually called ``derivative expansion''. 
For our mixed electromagnetic-gravitational case, 
higher derivative corrections to the effective action (\ref{drumhath}) due to a scalar loop were
considered in \cite{barvis}.
Those corrections can be summed up into
``Barvinsky-Vilkovisky form factors", which are closed-form integral expressions 
involving Schwinger-parameter type integrals. See the recent \cite{gusev} for the
state-of-the-art of this approach.

The second one corresponds to the standard heat-kernel or ``inverse mass'' expansion.
It is the most canonical one of the three in the sense that it is manifestly gauge and generally covariant 
order by order. The heat-kernel expansion of the one-loop effective action is usually written as

\bear
\Gamma[g,A] &=& \int_0^{\infty}{dT\over T} \,\e^{-m^2T}
\int {d^Dx\sqrt{g}\over (4\pi T)^{D\over 2}}
\sum_{n=0}^{\infty}a_n(x)T^n
\non\\
\label{defan}
\ear
where $D$ is the space-time dimension and $a_n(x)$ are the ``heat-kernel coefficients''.
In $D=4$ dimensions, the terms with $n=0,1,2$ are UV divergent at $T=0$, so that the
corresponding coefficients are subject to renormalization. 
For our case of the Einstein-Maxwell background with a spin 0 or spin 1/2 loop 
the coefficients can, up to $a_3$,
be obtained from more general results on the heat-kernel expansion \cite{gilkey,bafrts}. 
They are, for the scalar case
\footnote{To obtain the scalar loop coefficients from appendix B of \cite{bafrts}, replace
$E\to -\xi R$ and $F_{ab}\to i F_{ab}$. Here the parameter $\xi$ describes a non-minimal coupling to gravity. To obtain the spinor loop ones, replace
$E\to -\fourth R + {i\over 2}F_{ab}\gamma^a\gamma^b$ and
$F_{ab}\to \fourth R_{abcd}\gamma^c\gamma^d + i F_{ab}$.}

\bear
a_0 &=& 1 \, ,\non\\
a_1 &=& \Bigl(\sixth - \xi\Bigr) R \, ,\non\\
a_2 &=& -{1\over 12}F_{\mn}^2 \, ,\non\\
a_3 &=& {1\over 360}
\Bigl\lbrack
5(6\xi -1)RF_{\mn}^2 +4R_{\mn}F^{\mu\alpha}F^{\nu}{}_{\alpha}
-6R_{\mu\nu\alpha\beta}F^{\mn}F^{\ab} 
\non\\
&& \qquad
-2(\nabla^{\alpha}F_{\alpha\mu})^2
-8(\nabla_{\alpha}F_{\mn})^2
-12F_{\mn}\square F^{\mn}
\Bigr\rbrack
\non\\
\label{a2a3scal}
\ear
and for the spinor case,

\bear
a_0 &=& -2 \, ,\non\\
a_1 &=& \sixth R \, ,\non\\
a_2 &=& -{1\over 3}F_{\mn}^2 \, , \non\\
a_3 &=& {1\over 180}
\Bigl\lbrack
5RF_{\mn}^2 
-4R_{\mn}F^{\mu\alpha}F^{\nu}{}_{\alpha}
-9R_{\mu\nu\alpha\beta}F^{\mn}F^{\ab} 
\non\\
&& \qquad
+2(\nabla^{\alpha}F_{\alpha\mu})^2
-7(\nabla_{\alpha}F_{\mn})^2
-18F_{\mn}\square F^{\mn}
\Bigr\rbrack \, . \non\\
\label{a2a3spin}
\ear
Here the terms $a_0,a_1,a_2$ contribute to the renormalization of the vacuum energy,
Newton's constant, and electric charge, respectively. 

The expression (\ref{a2a3spin}) for the spinor case is equivalent to the one in 
(\ref{drumhath}), as can be seen by adding suitable total derivative terms; we will 
discuss this issue in section \ref{discussion} below. The scalar case result (\ref{a2a3scal})
in this explicit form is new, as far as we know. 

The third choice amounts to a generalization of the Euler-Heisenberg Lagrangian, the
object of interest in this paper.
Contrary to the pure QED case, for Einstein-Maxwell theory it is not obvious how one should
define the effective Lagrangian for constant external fields, since the notion of constancy
becomes ambiguous in curved space. 
To the best of our knowledge, the only previous attempt to treat the electromagnetic field 
and/or gravitational field nonperturbatively is due to Avramidi \cite{avramidi1, avramidi2}. 
This author generalizes the constancy of $F$ to the covariant constancy of $F$ and $R$,

\bear
\nabla_{\alpha}F_{\mn} &=& \nabla_{\alpha}R_{\mu\nu\kappa\lambda} =  0 \, .
\label{nablaF}
\ear
For a background obeying (\ref{nablaF}) he obtains an Euler-Heisenberg type formula
for the effective Lagrangian. However, the conditions (\ref{nablaF}) are rather strong,
and imply, for example, also a consistency condition between $F$ and $R$, since

\bear
\nabla_{\alpha}F_{\mn} &=& 0 \quad \rightarrow \quad 
\lbrack \nabla_{\alpha},\nabla_{\beta}\rbrack F_{\mn} = 0 
\quad \rightarrow \quad 
R_{\alpha\beta\mu\lambda}F^{\lambda}{}_{\nu}
-  R_{\alpha\beta\nu\lambda}F^{\lambda}{}_{\mu}
 = 0 \, .
\non\\
\label{conscond}
\ear
This strongly suggests that the effective action for this special case can carry only
some partial information on the low energy limit of the corresponding
amplitudes, i.e. the one-loop one particle irreducible (``1PI'') off-shell photon-graviton
amplitudes involving a scalar or spinor loop. In the present paper, we 
adopt a more general definition of a curved-space Euler-Heisenberg Lagrangian (``EHL'')
by demanding that, like the QED EHL, it should
contain the minimum set of terms in the covariant effective Lagrangian which would
have the full information on the low-energy limit of the corresponding 
1PI amplitudes.  
As usual in the graviton case, the amplitudes must be defined 
by linearizing gravity around flat Minkowski space. We will explicitly calculate the 
generalized EHL's for Einstein-Maxwell theory to linear order in the curvature, 
corresponding to the 1PI photon-graviton amplitudes with any number of photons but
not more than one graviton. It is easily seen that this truncation corresponds to keeping
all terms in the covariant effective Lagrangian which involve any number of
electromagnetic field strength tensors, together with up to one factor of the curvature tensor,
where this curvature tensor could also be replaced by two covariant derivatives.  
In this calculation, we use the recently completed extension of the worldline formalism 
\cite{polbook,berkos,strassler,ss1,review}      
to curved space \cite{basvanbook,baszir1,bacozi2,bacozi1,babegi} 
made manifestly covariant by using Riemann normal coordinates
and Fock-Schwinger gauge. 

The structure of this paper is as follows. In section
\ref{worldline} we summarize the worldline algorithm for
the calculation of one-loop effective actions in mixed gravitational-electromagnetic
fields. The calculation of the generalized Euler-Heisenberg Lagrangian
is presented in \ref{scalar} for the scalar and in \ref{spinor} for the spinor loop case.
In \ref{discussion} we compare with previous work and discuss possible applications
of these Lagrangians.
We summarize our findings in section \ref{conclusions}.
Our differential geometry conventions are given in appendix \ref{conventions}, where we
also collect some useful formulas.
In appendix \ref{green} we discuss some properties of the worldline Green's functions in
a constant field, to be introduced below.

\section{Worldline representation of the effective action in Einstein-Maxwell theory}
\label{worldline}
\renewcommand{\theequation}{2.\arabic{equation}}
\setcounter{equation}{0}

Let us start with the (euclidean) effective action for a complex scalar field $\phi$ coupled to
electromagnetism and gravity, 

\bea
S[\phi,\phi^* ;g,A] = -\int d^Dx \sqrt{g}
\Big [g^{\mu\nu} (\partial_\mu -i e A_\mu) \phi^* (\partial_\nu + ie A_\nu)
\phi + (m^2 +\xi R) \phi^* \phi \Big ]\non\\
\label{SgA}
\eea
where $\xi$ describes an additional non-minimal coupling to the scalar
curvature $R$.
Quantization produces the following effective action
($ Z[g,A] =e^{\Gamma[g,A]} =\int {\cal D}\phi {\cal D}\phi^*
\ e^{S[\phi,\phi^* ;g,A]} $)
\bea
\Gamma[g,A] = \ln {\det}^{-1} (-\square_A +m^2 +\xi R)
= - {\rm Tr} \ln  (-\square_A +m^2 +\xi R)
\label{2}
\eea
where $\square_A$ is the gauge and gravitational covariant laplacian for 
scalar fields. It  can  be represented by the following worldline path integral
(see, e.g., \cite{basvanbook,baszir1})
\footnote{Note that our definition of the global sign of the effective action 
follows \cite{review} rather than \cite{baszir1,phograv}.
It corresponds to a euclidean tree level action 
$\Gamma = -\int d^4x \sqrt{g}\, {1\over 4}F_{\mn}^2$.} 

\bea
\Gamma[g,A]
=  \int_0^\infty {dT\over T } \int_{PBC} {\cal D}x\ e^{-S[x^\mu;g,A]}
\label{Gammascal}
\eea
where
\bea
S[x^\mu;g,A] = \int_{0}^{1} d\tau \Big (
{1\over 4 T} g_{\mu\nu}(x) \dot x^\mu \dot x^\nu +
i e A_\mu(x) \dot x^\mu + T( \xi R(x) + m^2) \Big ) \ .
\non\\
\label{Lwl}
\eea
Here $T$ is the proper-time of the loop particle, and the path integral is to be
performed over all closed loops in spacetime $x(\tau)$ with periodic
boundary conditions $x(1) = x(0)$. 

Following the ``string-inspired'' procedure \cite{polbook,berkos,strassler,ss1} we will
evaluate the path integral $\int {\cal D}x(\tau)$ 
by manipulating it into gaussian form, using a double expansion.
First, one Taylor expands the external fields at some point $x_0$
\cite{ss1,fss1,fhss4,gussho,dilmck}.
This is most conveniently done in covariant form, i.e., using a combination of
Fock-Schwinger gauge and Riemann normal coordinates \cite{alfrmu,dilmck}:

\bear
g_{\mu \nu}(x=x_0+y)&=&g_{\mu \nu}(x_{0})+\frac{1}{3}R_{\mu \alpha \beta \nu}(x_{0})\,y^{\alpha}\,y^{\beta}
+\cdots \nonumber\\
\label{Riemannexp}
\ear
\bear\label{A des}
A_{\mu}(x=x_0+y)&=&-\frac{1}{2}F_{\mu \nu}(x_{0})\,y^{\nu}-\frac{1}{3}F_{\mu \nu ; \alpha}(x_{0})\,y^{\nu}\,y^{\alpha}-\frac{1}{8}\bigg[ F_{\mu \nu ; \alpha \beta}(x_{0}) \nonumber\\
&& +\frac{1}{3}R_{\alpha\mu}{}^{\lambda}{}_\beta (x_0)F_{\lambda\nu}(x_{0}) \bigg]y^{\alpha}\,y^{\beta}\,y^{\nu}+\cdots 
\label{FS}
\ear
(see appendix \ref{conventions} for
our Riemannian geometry conventions).
The worldline action then takes the form 

\bear
S[x^{\mu};R,F] \, & =& {1\over 4T} \int_0^1 d\tau
\,\dot y^{\mu}(\tau)g_{\mn}(x_0)\dot y^{\nu}(\tau) 
 + S_{\rm int}[x^{\mu};R,F]
 \non\\
\label{Snew}
\ear
where $S_{\rm int}[x^{\mu};R,F]$ contains an infinite number of interaction terms.  
In principle, all these interaction exponentials are to be expanded out, although 
realistically the arising multiple series has to be truncated to some desired level.
Usually this truncation will be
either in the number of fields, in the number of derivatives, or in the 
mass dimensions. 

Next, one has to fix the zero mode of the path integral, due to translation invariance
in spacetime. There are two standard ways of doing this, both using a restriction of the
path integration to fluctuations around the expansion point $x_0$,

\be
x^\mu(\tau)= x_0^\mu+y^\mu(\tau),
\label{defy}
\ee
where  the  path  integral measure is split into
\bea
 Dx= {d^Dx_0\sqrt{g(x_0)}\over (4 \pi T)^{D \over 2}}\, 
 Dy \ .
\label{splitmeasure}
\eea
The choice is in the constraints imposed on $y(\tau)$, which are either
Dirichlet boundary conditions

\bear
y(0) = y(1) = 0 
\label{DBC}
\ear
meaning that $x_0$ is on the loop
(``DBC scheme''),
or the ``string-inspired'' condition (``SI scheme'')

\bear
\int_0^1 d\tau \, y^{\mu}(\tau) &=& 0
\label{defSI}
\ear
which makes $x_0$ the center of mass of the loop, see figures 1 and 2.

\begin{figure}[t]
\begin{minipage}[b]{0.5\linewidth}
\centering
\includegraphics[scale=.8]{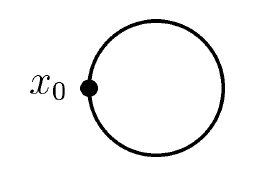}
\caption{DBC scheme}
\label{fig:DBC}
\end{minipage}
\hspace{0cm}
\begin{minipage}[b]{0.5\linewidth}
\centering
\includegraphics[scale=.8]{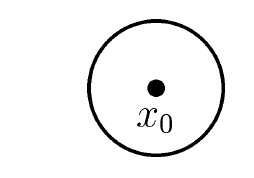}
\caption{SI scheme}
\label{fig:SI scheme}
\end{minipage}
\end{figure}

The DBC scheme leads to a worldline propagator

\bea
\langle y^\mu (\tau) y^\nu(\sigma)\rangle \!\! &=& \!\!
- 2 T g^{\mu\nu}(x_0)  
\Delta(\tau,\sigma)
\label{wickDBC}
\eea
where

\bear
\Delta (\tau,\sigma)  &=&
 \sum_{m=1}^{\infty}
\biggl [ - {2 \over {\pi^2 m^2}} \sin (\pi m \tau)
\sin (\pi m \sigma) \biggr ] 
\non\\
&=&
(\tau-1)\sigma\, \theta(\tau-\sigma)+(\sigma-1)\tau\, \theta(\sigma-\tau) \, .
\non\\
\label{defDelta}
\eea
The propagator in the SI scheme is

\bea
\langle y^\mu (\tau) y^\nu(\sigma)\rangle \!\! &=& \!\!
- T g^{\mu\nu}(x_0)  
G_B(\tau,\sigma)
\label{wickSI}
\eea
where

\bear
G_B(\tau,\sigma) &=&
2\sum_{m=-\infty\atop m\ne 0}^{\infty}
{\e^{2i\pi m (\tau -\sigma)}\over (2i\pi m)^2} =
\vert \tau -\sigma\vert -\Bigl(\tau -\sigma\Bigr)^2-{1\over 6}\, .\non\\
\label{GB}
\ear
For gauge theory in flat space, either propagator can be chosen for a straightforward 
perturbative calculation of the one-loop effective action via formal gaussian integration
\footnote{In flat space calculations the constant $-{1\over 6}$ in $G_B$ does not affect physical
quantities and is therefore usually omitted.}.
The effective Lagrangian obtained in the DBC scheme 
coincides with the heat kernel result, while the SI scheme differs from it,
but only by total derivative terms \cite{fss1,fhss4,review}.
Thus both schemes are completely equivalent, but the SI scheme is computationally
preferable, since it preserves the translation invariance in the proper-time.

Proceeding to the inclusion of gravitational backgrounds, here 
a number of mathematical difficulties arise which are not present in flat space,
starting with the observation that the structure of the worldline Lagrangian (\ref{Lwl})
generically leads to ill-defined expressions involving, e.g., 
$\delta(0), (\delta(\tau-\sigma))^2, \ldots$ in a gaussian integration. 
A completely satisfactory formalism for dealing with these issues has emerged only
in recent years \cite{bastianelli,basvan93,depeskva,bascva,kleche,bacova}. 
Here we can only sketch the procedure; a brief discussion appears in \cite{bast} and
all the details can
be found in \cite{basvanbook}. 

First, in curved space the path integral measure is nontrivial. Following
\cite{bastianelli,basvan93} we exponentiate it as follows,

\bear
{\cal D} x &=& Dx \prod_{ 0 \leq \tau < 1} \sqrt{\det g_{\mu\nu}(x(\tau))} 
=
Dx \int_{PBC} { D} a { D} b { D} c \;
{\rm e}^{- S_{gh}[x,a,b,c]}, \non\\
 \label{expmeasure}
\ear
with a ghost action 

\bear
S_{gh}[x,a,b,c]
= \int_{0}^1 d\tau \; {1\over 4T}g_{\mu\nu}(x)\Bigl\lbrack a^\mu (\tau) a^\nu (\tau) 
+ b^\mu (\tau) c^\nu (\tau) \Bigr\rbrack \, .\label{Sghost}
\ear
After the replacement of $g_\mn (x)$ by its normal coordinate expansion
(\ref{Riemannexp}) the correlators of these ghost fields just involve 
$\delta$ functions,

\bear
\langle a^{\mu}(\tau)a^{\nu}(\sigma)\rangle
&=&
2T g^{\mn}(x_0)\delta(\tau-\sigma), \non\\
\langle b^{\mu}(\tau)c^{\nu}(\sigma)\rangle
&=&
-4T g^{\mn}(x_0)\delta(\tau-\sigma) \, .\non\\
\label{wickghost}
\ear
The ghost field contributions will cancel all ill-defined divergent terms
of the type mentioned above, arising from the Wick contractions of the coordinate
fields. 
This cancellation of infinities generally still leaves integrals with 
ambiguities. A basic example is

\bear
\int_0^1d\tau \int_0^1 d\sigma \,\delta(\tau-\sigma)
\theta(\tau -\sigma)\theta(\sigma -\tau) \, .
\label{example}
\ear
This type of integral requires a regularization, and different regularizations will
assign different finite values to it \cite{basvanbook}.

From the point of view of one-dimensional quantum field theory, we are dealing
here with a theory which, without the terms from the nontrivial path integral measure, 
would be UV divergent but super-renormalizable. Including those terms removes all
divergences, but leaves finite ambiguities, so that agreement with standard 
spacetime QFT is reached only after adding a finite number of counterterms with finite,
regularization-dependent  coefficients.
The method which we will adopt here is 
(one-dimensional) dimensional regularization \cite{kleche,bacova}, since it is presently
the only known regulator which preserves the general covariance. 
It needs only a single counterterm proportional to the curvature scalar,
$-R/4$ in the present notations. Therefore in this scheme 
the only effect of the spurious UV divergences is a
change of the parameter $\xi$ in the worldline Lagrangian (\ref{Lwl})
into ${\bar \xi}$,

\bear
{\bar \xi} := \xi - \fourth\, .
\label{xixibar}
\ear
A further subtlety shows up if one wishes to combine the SI scheme with the
use of the Riemann-Fock-Schwinger expansion (\ref{Riemannexp}),(\ref{FS}). 
As is well-known, this expansion is useful only for the calculation of
covariant quantities; when applied to non-covariant quantities, it yields
a result which is formally covariant but correct only in Riemann normal coordinates.
The DBC scheme in curved space still
yields the same effective Lagrangian as the standard heat-kernel
method, and thus also guarantees covariance. This does not extend to the
SI scheme, since it turns out that the total derivative terms by which the SI effective
Lagrangian differs from the DBC one generally are not covariant \cite{schvan,baszir1}. 
A solution to this problem convenient for actual calculations was found in \cite{bacozi2}. 
There it was shown that, using Riemann normal coordinates
from the beginning and performing a BRST treatment of the symmetry corresponding to a 
shift of $x_0$, the difference between the two effective Lagrangians can be reduced
to manifestly covariant terms. This is achieved by the addition of further Fadeev-Popov type 
terms to the worldline Lagrangian in the ``string-inspired'' scheme.
Those terms are infinite in number but easy to determine order by order.
In our present approximation, this Fadeev-Popov action can be truncated as

\bea
S_{FP} = - \bar\eta_\mu   \int_0^1 d\tau\,  
Q^\mu_\nu(x_0,y(\tau)) \, \eta^\nu
\eea
where 
\bea
Q^\mu{}_\nu(x_0,y)&=&
\delta^\mu_\nu + 
{1\over 3} R^\mu{}_{\alpha\beta\nu}(x_0)\, y^\alpha y^\beta 
+ ....
\eea
The propagator for the (constant) ghost fields $\eta, \bar\eta$ is trivial,

\bea
\langle \eta^\mu \bar \eta_\nu \rangle 
&=& - \delta^\mu_\nu      \ .
\label{propeta}
\eea

Having concluded our discussion of the scalar loop case, we proceed to
the case of a spin 1/2 particle in the loop.
The euclidean action for a Dirac field $\Psi$ coupled to
electromagnetism ($A_\mu$) and gravity ($e_\mu{}^a$) is given by
\bea
S[\Psi,\bar \Psi;e, A]
= -\int \!d^Dx \; e\, \bar{\Psi} (\rldd + m)\Psi  
\eea
where $e_\mu{}^a$ is the vielbein,
$ e= \det e_\mu{}^a$, $\omega_{\mu ab} $ is the spin connection, and
\be
\rldd = \gamma^a e_a{}^\mu \nabla_{\!\mu} \ , \quad \quad 
\nabla_{\!\mu} = \partial_\mu + i e A_\mu 
+ {1\over 4}\omega_{\mu ab} \gamma^a \gamma^b \ .
\ee
The effective action depends on the background fields $e_\mu{}^a$ and $A_\mu$,
and formally reads as  
($ e^{\Gamma[e,A]} \equiv \int {\cal D}\Psi\, {\cal D}\bar \Psi 
\ e^{S[\Psi,\bar\Psi; e,A]} = {\rm Det} (\rldd + m)$) 
\bea
\Gamma[e,A] 
\eqa
\ln {\rm Det} (\rldd + m) 
= \ln [{\rm Det} (\rldd + m) {\rm Det} (-\rldd + m)]^{1\over 2} 
\ccr
\eqa
{1\over 2} {\rm Tr} \ln  (-\rldd^2 + m^2) 
\ccr
\eqa
 {1\over 2} {\rm Tr} \ln  \Big (-\square_A +m^2+ {1\over 4}R
\Big ) 
\ .
\eea

A worldline path integral representation for this effective action can be 
written in a manifestly local Lorentz invariant way \cite{bacozi1}
(i.e. in terms of the metric rather than the vielbein)
\bea
\Gamma[g, A] 
= -{1\over 2} \int_0^\infty {dT\over T } \int_{PBC} \!\!\!\!\!\!\!
{\cal D}x
\int_{ABC} \!\!\!\!\!\!\!
{\cal D}\psi \; e^{-S[x^\mu,\psi^\mu; g,A]}                 
\label{Gammaspin}
\eea
with
\bea
S[x^\mu,\psi^\mu; g,A]  \eqa
\int_{0}^{1} \! \! d\tau\, \bigg [
{1\over 4 T} g_{\mu\nu}(x) \dot x^\mu \dot x^\nu + i e A_\mu(x) \dot x^\mu 
+T \bigl(\fourth R + m^2\bigr) \ccr
&&  \quad  +
{1\over 2 T} \Bigl(g_{\mu\nu}(x) \psi^\mu \dot{\psi}^\nu
- \partial_\mu g_{\nu\lambda}(x) \psi^\mu\psi^\nu \dot x^{\lambda} \Bigr)
- i e F_{\mu\nu}(x)\psi^\mu \psi^\nu
\bigg ] \, .
\non\\
\label{Lwlspin}
\eea
Note that the bosonic term appearing in the first line is the same as for 
a scalar particle with $\xi=\fourth$, see eq. (\ref{Lwl}) and (\ref{xixibar}), 
while the second term contains the worldline fermions and describes the dependence on the spin 
of the particle.
This action also makes it clear that there are only 
linear couplings of the spin 1/2 particle to the metric $g_{\mu\nu}$.
The worldline fields $\psi^{\mu}(\tau)$ are Grassmann valued and antiperiodic,
$\psi(1)=-\psi(0)$. The free spin path integral is normalized as $2^{D/2}$.
(Note also that our $\psi^{\mu}$ corresponds to $\sqrt{T}\psi^{\mu}$ 
in the conventions of \cite{review}.)

Again we gaussianize the double path integral in (\ref{Gammaspin}) by the
use of the Riemann-Fock-Schwinger expansion (\ref{Riemannexp}),(\ref{FS}). Now also
the expansion of $F_{\mn}$ is needed, which follows from (\ref{FS}):

\bear
F_{\mu \nu}(x_0+y)&=&F_{\mu \nu}(x_0)+F_{\mu \nu ; \alpha}(x_0)\,y^{\alpha}+\frac{1}{2}F_{\mu \nu ; \alpha \beta}(x_0)\,y^{\alpha}\,y^{\beta} \nonumber\\
&&+\frac{1}{6} \Big( 
R_{\alpha\mu}{}^{\lambda}{}_\beta (x_0) F_{\lambda\nu}(x_0)
+
R_{\alpha\nu}{}^{\lambda}{}_\beta (x_0) F_{\mu\lambda}(x_0)
\Big)\,y^{\alpha}\,y^{\beta}+\ldots \nonumber\\
\label{Fexpand}
\ear
Eq. (\ref{Snew}) generalizes to

\bear
S[x^{\mu},\psi^{\nu};R,F] \, & =& {1\over T} \int_0^1 d\tau
\,\Bigl\lbrack\fourth\dot y^{\mu}(\tau)g_{\mn}(x_0)\dot y^{\nu}(\tau) 
+\half \psi^{\mu}(\tau)g_{\mn}(x_0)\dot\psi^{\nu}(\tau) \Bigr\rbrack
\non\\[2mm]
&& + S_{\rm int}[x^{\mu},\psi^{\nu};R,F] \, .
 \non\\
\label{Snewspin}
\ear
The propagator of the worldline fermions then becomes

\bea
\langle \psi^\mu (\tau) \psi^\nu(\sigma)\rangle \!\! &=& \!\!
\half T g^{\mu\nu}(x_0)  
G_F(\tau,\sigma)
\label{wickSIGF}
\eea
where

\bear
G_F(\tau,\sigma) &=&
2\sum_{m=-\infty}^{\infty}
{\e^{i\pi (2m+1) (\tau -\sigma)}\over i\pi (2m+1)} =
{\rm sign}(\tau -\sigma) \, .
\non\\
\label{GF}
\ear
Note that, due to the antisymmetry of the spin path integral, there is no zero mode and
thus no related ambiguity for this propagator.

Like the bosonic path integral measure, the fermionic one is nontrivial in curved space,
leading to a generalization of (\ref{expmeasure}) to

\bear
{\cal D} x {\cal D} \psi
&=&
Dx D\psi \int_{PBC} { D} a { D} b { D} c \int_{ABC}{ D} \alpha\;
{\rm e}^{- S_{gh}[x,a,b,c,\alpha]}, \non\\
 \label{expmeasurespin}
\ear
where the ghost action now is

\bear
S_{gh}[x,a,b,c,\alpha]
= \int_{0}^{1} d\tau \; {1\over 4T}g_{\mu\nu}(x)\Bigl\lbrack a^\mu (\tau) a^\nu (\tau) 
+ b^\mu (\tau) c^\nu (\tau) +2 \alpha^\mu (\tau) \alpha^\nu (\tau)\Bigr\rbrack\, .
 \non\\
\label{Sghostspin}
\ear
The correlator of the new ghost field $\alpha^\mu (\tau)$ is,
after the normal coordinate expansion,

\bear
\langle \alpha^{\mu}(\tau)\alpha^{\nu}(\sigma)\rangle
&=&
T g^{\mn}(x_0)\delta(\tau-\sigma)  \, .
\label{wickghostspin}
\ear
Again there are cancellations of ill-defined divergent terms between the $\psi$ and the $\alpha$
path integrals, forcing one to choose a regularization and possibly leading to
a modification of the counterterms introduced for the spinless worldline Lagrangian
above. 
However, it turns out that in dimensional regularization this does not happen; the 
sole counterterm $-\fourth R$ remains also the correct one for the spin 1/2 case \cite{bacozi1}.
Its effect is just to remove the term linear in $R$ which was there in the
initial worldline Lagrangian (\ref{Lwlspin}). 
Other regularizations have been discussed in \cite{bonfal}.

In principle, this is all one has to know for calculating the one-loop effective action
for spin 0 or spin 1/2 particle in the Einstein-Maxwell background, or the corresponding
amplitudes \cite{baszir1,bacozi1}. 
However, since we are aiming at a result which is nonperturbative in the
electromagnetic field, for us the following modification will be essential:
Note that the leading terms of the Fock-Schwinger expansions (\ref{FS}) and
(\ref{Fexpand}) yield terms in $S_{\rm int}$ which are quadratic in the worldline fields.
Thus instead of using them in the interaction part one can absorb them in the
worldline propagators. In the SI scheme this leads to the
following change of the correlators (\ref{wickSI}), (\ref{wickSIGF})
\cite{shaisultanov,rescsc,vv}, 

\bear
\langle y^\mu (\tau) y^\nu (\sigma) \rangle &=& -T{\cal G}_B^\mn (\tau,\sigma) \, , \non\\
\langle \psi^\mu (\tau) \psi^\nu (\sigma)\rangle &=& \half T{\cal G}_F^\mn (\tau,\sigma) \, . \non\\
\label{wickF}
\ear
The new worldline propagators are trigonometric
functions of the field strength matrix $F$,
and thus, in general, nontrivial Lorentz matrices:

\bear
{\cal G}_B^\mn (\tau_1,\tau_2) &=&
\biggl\lbrack
{1\over 2{(FT)}^2}\Biggl({FT\over{{\rm sin}(FT)}}
{\rm e}^{-iFT\dot G_{B12}}
\!+\! iFT\dot G_{B12}\! - 1\Biggr)
\biggr\rbrack^\mn \, ,
 \non\\
{\cal G}_F^\mn (\tau_1,\tau_2) &=&
\biggl\lbrack
G_{F12}
{{\rm e}^{-iFT\dot G_{B12}}\over \cos (FT)}
\biggr\rbrack^\mn \, .
 \non\\
\label{calGBGF}
\ear
Here and in the following we abbreviate $G_{B12}=G_B(\tau_1,\tau_2)$ etc., and 
a `dot' on a Green's function denotes a derivative with respect to the first variable.
In the computation of the power series 
appearing in the definitions (\ref{calGBGF}) it 
should be understood that indices are raised and lowered with the metric $g_{\mn}(x_0)$.

Finally, the free gaussian path integrals get also modified and become field-dependent; namely, the
coordinate path integral acquires a  factor of 
${\rm det}^{-\half}\lbrack \sin (FT) / FT\rbrack$,
the spin path integral a 
${\rm det}^{\half} \lbrack \cos (FT) \rbrack$.
By themselves these factors just reproduce the integrands of the 
(unrenormalized) Euler-Heisenberg
Lagrangians (\ref{ehspin}), (\ref{ehscal}). 

This version of the formalism has already been applied extensively to the calculation of   
pure QED amplitudes or effective actions in a constant background field 
\cite{adlsch,rescsc,frss,korsch,gussho}. 
More recently
it has been used for a first calculation of the photon-graviton polarization tensor in
a constant field \cite{phograv}. See also \cite{rescsc,satsch2,sascza,pasazh}
for an extension to the nonabelian case.

\section{Calculation of the effective Lagrangian: scalar loop}
\label{scalar}
\renewcommand{\theequation}{3.\arabic{equation}}
\setcounter{equation}{0}

In the following we specialize to the SI scheme. 
According to the above, the worldline representation of the effective Lagrangian
for the scalar loop in $D=4$ dimensions in this scheme can be written as
\footnote{In the following it is understood that all spacetime fields
are sitting at the expansion point $x_0$, and $\Gamma = \int d^4x_0 \sqrt{g}{\cal L}$.}

\bear
{\cal L}_{\rm scal} &=&
{1\over 16\pi^2}
\int_0^{\infty}{dT\over T^3}
\,\e^{-m^2T}
{\rm det}^{-\half}\Bigl\lbrack {\sin (FT) \over FT}\Bigr\rbrack
\Bigl\langle \,\e^{-S_{\rm int}[x^\mu,a,b,c,\eta;R,F]}\Bigr\rangle \, .
\non\\
\label{Lscalcap3}
\ear
In our one-graviton approximation, the worldline interaction Lagrangian can be
truncated as

\bear
S_{\rm int} &=& S_{grav}+S_{em}+S_{em,grav}+S_{gh}+S_{FP}\, ,
\label{Sintscal}
\ear

		\bear\label{Sgravscal}  
 		  S_{grav}+S_{gh}&=&
T\bar{\xi}\,R+\frac{1}{12T}\int ^{1}_{0}\,d\tau R_{\mu \alpha \beta \nu} y^{\alpha}y^{\beta}\biggl\lbrack
\dot{y}^{\mu}\dot{y}^{\nu}+a^{\mu}a^{\nu}+b^{\mu}c^{\nu}\biggr\rbrack,\nonumber\\
		\ear

		\bear
                  S_{em}&=&\int ^{1}_{0}d\tau \bigg[
-\frac{i}{3} F_{\mu \nu ; \alpha} \dot{y}^{\mu} y^{\nu} y^{\alpha}
                 -\frac{i}{8}  F_{\mu \nu; \alpha \beta} \,\dot{y}^{\mu}\, y^{\nu}\,y^{\alpha}\,y^{\beta}\,\bigg], \non\\
		\ear

\bear
		  S_{em,grav}&=&-\frac{i}{24}\int ^{1}_{0}d\tau 
R_{\alpha\mu}{}^{\lambda}{}_{\beta}F_{\lambda\nu}
y^{\nu}\,y^{\alpha}\,y^{\beta}\,\dot{y}^{\mu} , \nonumber\\
		\ear

\begin{equation} \label{Sfp}
S_{FP}=-\third \int ^{1}_{0}d\tau \ \bar{\eta}_\mu 
R^{\mu}{}_{\alpha \beta \nu}\,y^\alpha y^\beta
\, \eta^\nu.
\end{equation}
Note that the term involving $y^\mu F_{\mn}\dot y^\nu$ has been omitted from $S_{\rm em}$.

For easy reference, let us also list the complete set of worldline propagators
of the SI scheme:

\bear
\langle y^\mu(\tau)  y^\nu(\sigma)\rangle &=& 
- T {\cal G}_B^\mn(\tau,\sigma) \, , \non\\
\langle a^\mu (\tau) a^{\nu} (\sigma) \rangle &=& 2Tg^\mn \delta(\tau-\sigma) \, ,\non\\
\langle b^\mu (\tau) c^{\nu} (\sigma) \rangle &=& -4Tg^\mn \delta(\tau-\sigma) \, ,\non\\
\langle \eta^\mu \bar\eta_\nu \rangle &=& -\delta^{\mu}_{\nu} \, ,\non\\
\label{SIproplistscal}
\ear
where ${\cal G}_B$ was given in (\ref{calGBGF}).

With all this machinery in place, it is then straightforward to obtain the following result
for the (unrenormalized) scalar loop effective Lagrangian in the one-graviton approximation,

\begin{eqnarray}
{\cal L}_{\rm scal}^{(SI)}&=&{1\over 16\pi^2}
\int^{\infty}_{0} \frac{dT}{T^3}\,\e^{-m^2T}
\mbox{det}^{-1/2}\left[ \frac{\sin(FT)}{FT}\right]\Biggl\lbrace 1-T\bar{\xi}R 
+\frac{T}{3}{\cal G}^{\alpha \beta}_{B11}R_{\alpha \beta}\nonumber\\
&&  +\frac{iT^2}{8}F_{\mu \nu ;  \alpha \beta}\,\dot{{\cal G}}^{\mu \nu}_{B11}\,{\cal G}^{\alpha \beta}_{B11}+\frac{i}{8}T^2\left(F_{\mu \nu ; \beta \alpha} + F_{\mu \nu ;  \alpha \beta}\right)\dot{{\cal G}}^{\mu \beta}_{B11}{\cal G}^{\nu \alpha}_{B11} \nonumber\\
&&-\frac{iT^2}{24}F_{\lambda \nu}R^{\lambda}_{\, \, \, \alpha \beta\mu}\,\left(\dot{{\cal G}}^{\nu \mu}_{B11}\,{\cal G}^{\alpha \beta}_{B11}+\dot{{\cal G}}^{\alpha \mu}_{B11}\,{\cal G}^{\nu \beta}_{B11}+\dot{{\cal G}}^{\beta \mu}_{B11}\,{\cal G}^{\nu \alpha}_{B11}\right) \nonumber\\
&&+\frac{T}{12}R_{\mu \alpha \beta \nu}\left( \dot{{\cal G}}^{\mu \alpha}_{B11}\dot{{\cal G}}^{\beta \nu}_{B11}+\dot{{\cal G}}^{\mu \beta}_{B11}\dot{{\cal G}}^{\alpha\nu}_{B11}
+\Bigl(\ddot{{\cal G}}^{\mu \nu}_{B11}-2g^\mn \delta(0)\Bigr){\cal G}^{\alpha \beta}_{B11}
\right) 
\nonumber\\
&&-\frac{T^3}{6}F_{\alpha \beta;\gamma}F_{\mu \nu ;\delta}
\int^{1}_{0}d\tau_1\left(\dot{{\cal G}}^{\alpha \nu}_{B12}\, \dot{{\cal G}}^{\beta \mu}_{B12} \, {\cal G}^{\gamma \delta}_{B12}+\dot{{\cal G}}^{\alpha \nu}_{12}\, {\cal G}^{\beta \delta}_{12} \, \dot{{\cal G}}^{\gamma \mu}_{B12}\right) 
\Biggr\rbrace. \non\\
\label{resultscal}
		\end{eqnarray}
Here in the last term it is understood that $\tau_2 = 0$. Although 
getting (\ref{resultscal}) from (\ref{Lscalcap3}) is a matter of standard combinatorics,
a few comments are in order:

\benn

\item
In the next-to-last term in braces the $\delta(0)$ comes from 
the ghost sector and substracts a $\delta(0)$ contained in $\ddot {\cal G}_{B11}$.

\item 
In the last term in braces the Wick contractions produce also terms involving
a contracting among the fields inside one factor of 
$F_{\mu \nu ; \alpha} \dot{y}^{\mu} y^{\nu} y^{\alpha}$,
however those have vanishing $\tau_1$ or $\tau_2$ integrals due to (\ref{intort}).
The remaining terms have been reduced to a minimal set using integrations
by parts in $\tau_1$ and the Bianchi identity (\ref{bianchiF}). 

\item
The third term in braces comes from the Fadeev-Popov part $S_{FP}$ of the
worldline action. The inclusion of this term is necessary to obtain the equivalence with the standard
heat kernel result, to be shown in section \ref{discussion} below. This confirms
the formal reasoning of \cite{bacozi2}.

\item
No ambiguous integrals are encountered yet at the level of our
present calculation, so that regularization was not really necessary.
This can be understood from the fact that any arising ambiguity would
have to be cancelled by regularization dependent counterterms.
Those generally involve products of Christoffel symbols \cite{basvanbook}, 
and can therefore
in Riemann normal coordinates appear only starting at the quadratic level
in the curvature.

\enn

\section{Calculation of the effective Lagrangian: spinor loop}
\label{spinor}
\renewcommand{\theequation}{4.\arabic{equation}}
\setcounter{equation}{0}

For the spinor loop, the analogue of (\ref{Lscalcap3}) is

\bear
{\cal L}_{\rm spin} &=&
-{1\over 8\pi^2}
\int_0^{\infty}{dT\over T^3}
\,\e^{-m^2T}
{\rm det}^{-\half}\Bigl\lbrack {\tan (FT) \over FT}\Bigr\rbrack
\Bigl\langle \,\e^{-S_{\rm int}[x^\mu,\psi^\mu,a,b,c,\alpha,\eta;R,F]}\Bigr\rangle
\, .
\non\\
\label{Lspincap4}
\ear
The various components of the worldline interaction Lagrangian (\ref{Sintscal}) 
generalize as follows:

\begin{eqnarray}\label{Sgravspin}  
S_{grav}+S_{gh}&=&{1\over T}
\int ^{1}_{0}\,d\tau\Biggl\lbrace \frac{1}{12} R_{\mu \alpha \beta \nu} y^{\alpha}y^{\beta}
\biggl\lbrack
\dot{y}^{\mu}\dot{y}^{\nu}+a^\mu a^\nu + b^\mu c^\nu +2\alpha^\mu \alpha^\nu
\biggr\rbrack
\non\\&&
+\frac{1}{6}R_{\mu \alpha \beta \nu}\,y^{\alpha}\,y^{\beta}\,\psi^{\mu}\,\dot{\psi}^{\nu}
+\frac{1}{6}( R_{\mu  \alpha \lambda \beta}+R_{\mu \beta \lambda \alpha} ) \dot{y}^{\alpha}\,y^{\lambda}\,\psi^{\mu}\, \psi^{\beta}\Biggr\rbrace,\non\\
\end{eqnarray}

\begin{eqnarray}  
 S_{em}&=&\int ^{1}_{0}d\tau \bigg[ -\frac{i}{3}F_{\mu \nu ; \alpha} \Big( \dot{y}^{\mu}\, y^{\nu}+3\psi^{\mu}\,\psi^{\nu}\Big)y^{\alpha} -\frac{i}{8} F_{\mu \nu; \alpha \beta}  \Big( \dot{y}^{\mu}\,y^{\nu}\,+4\psi^{\mu}\psi^{\nu} \Big)\,y^{\alpha}\,y^{\beta}\bigg], \nonumber\\
\end{eqnarray} 

\begin{eqnarray}
S_{em,grav}&=&-\frac{i}{24}\int ^{1}_{0}d\tau
R_{\alpha\mu}{}^{\lambda}{}_{\beta}F_{\lambda\nu}
\Bigl[ \dot{y}^{\mu}\, y^{\nu} +8 \psi^{\mu}\psi^{\nu} \Bigr]\,y^{\alpha}\,y^{\beta} \, .\non\\
\end{eqnarray}
$S_{FP}$ is not modified from its form for the spinless case, eq. (\ref{Sfp}).
In addition to the propagators of the scalar case, (\ref{SIproplistscal}),
we have now also 

\bear
\langle \psi^\mu (\tau) \psi^\nu(\sigma)\rangle &=& 
\half T{\cal G}_F^\mn(\tau,\sigma) \, , \non\\
\langle \alpha^\mu (\tau) \alpha^{\nu} (\sigma) \rangle &=& Tg^\mn \delta(\tau-\sigma) \, .\non\\
\label{SIproplistspin}
\ear
The final result for the spinor loop case becomes

\begin{eqnarray}
{\cal L}_{\rm spin}^{(SI)}&=&
-{1\over 8\pi^2}
\int^{\infty}_{0} \frac{dT}{T^3}\,\e^{-m^2T}\mbox{det}^{-1/2}\left[ \frac{\tan(FT)}{FT}\right] 
\nonumber\\&&\times
\Biggl\lbrace
1+\frac{iT^2}{8}F_{\mu \nu ;  \alpha \beta}\,\,{\cal G}^{\alpha \beta}_{B11}\Big(\dot{{\cal G}}^{\mu \nu}_{B11}-2\,{\cal G}^{\mu \nu}_{F11} \Big) \nonumber\\
&&+\frac{i T^2}{8}\left(F_{\mu \nu ; \beta \alpha} + F_{\mu \nu ;  \alpha \beta}\right)\dot{{\cal G}}^{\mu \beta}_{B11}{\cal G}^{\nu \alpha}_{B11}+\frac{T}{3}R_{\alpha \beta}\,{\cal G}^{\alpha \beta}_{B11} \nonumber\\
&&-\frac{i T^2}{24}F_{\lambda \nu}R^{\lambda}_{\, \, \, \alpha \beta\mu}\,\left(\dot{{\cal G}}^{\nu \mu}_{B11}\,{\cal G}^{\alpha \beta}_{B11}+\dot{{\cal G}}^{\alpha \mu}_{B11}\,{\cal G}^{\nu \beta}_{B11}+\dot{{\cal G}}^{\beta \mu}_{B11}\,{\cal G}^{\nu \alpha}_{B11}+4\,{\cal G}^{\mu \nu}_{F11}\,{\cal G}^{\alpha \beta}_{B11}\right) \nonumber\\
&&+\frac{T}{12}R_{\mu \alpha \beta \nu}\Big(\dot{{\cal G}}^{\mu \alpha}_{B11}\dot{{\cal G}}^{\beta \nu}_{B11}+\dot{{\cal G}}^{\mu \beta}_{B11}\dot{{\cal G}}^{\alpha\nu}_{B11}
+\Bigl(\ddot{{\cal G}}^{\mu \nu}_{B11}-2g^\mn\delta(0)\Bigr){\cal G}^{\alpha \beta}_{B11}
\non\\
&&+\dot{{\cal G}}^{\alpha \beta}_{B11}\,{\cal G}^{\mu \nu}_{F11}
+\dot{{\cal G}}^{\nu \beta}_{B11}\,{\cal G}^{\mu \alpha}_{F11}
-{\cal G}^{\alpha \beta}_{B11}\,\Bigl(\dot{{\cal G}}^{\mu \nu}_{F11}-2g^\mn\delta(0)\Bigr)
\Big) \nonumber\\
&&-\frac{1}{6}T^{3}F_{\alpha \beta; \gamma}\,F_{\mu \nu ; \delta}\,\int^{1}_{0}d\tau_{1}\Big(\dot{{\cal G}}^{\alpha \nu}_{B12}\,\dot{{\cal G}}^{\beta \mu}_{B12} \, {\cal G}^{\gamma \delta}_{B12}+\dot{{\cal G}}^{\alpha \nu}_{B12}\,{\cal G}^{\beta \delta}_{B12} \, \dot{{\cal G}}^{\gamma \mu}_{B12} \nonumber\\
&&+\frac{3}{2}\,{\cal G}^{\gamma \delta}_{B12}\,{\cal G}^{\alpha \mu}_{F12}\,{\cal G}^{\beta \nu}_{F12}
\Big)  \Biggr\rbrace     \nonumber\\
\label{resultspin}		
\end{eqnarray}
where again $\tau_2 = 0$.

\section{Comparison with previous results}
\label{discussion}
\renewcommand{\theequation}{5.\arabic{equation}}
\setcounter{equation}{0}

As a check on our effective Lagrangians (\ref{resultscal}), (\ref{resultspin}), let us
extract the terms corresponding to the heat kernel coefficients $a_3$. This can be easily done
using formulas (\ref{GB(F)expand}), and yields, after performing the global proper-time integration,

\bear
{\cal L}_{\rm scal}^{(SI)} &=& {1\over 16\pi^2} \frac{e^2}{m^2}
\biggl\lbrack
\frac{1}{12}(\bar\xi + \frac{1}{12}) R F_\mn^2 
+ \frac{1}{180}R_\mn F^{\mu\alpha}F^{\nu}{}_\alpha \non\\
&&
- \frac{1}{72}R_{\mu\nu\alpha\beta}F^\mn F^\ab
- \frac{1}{180}(\nabla_{\alpha}F_{\mn})^2
- \frac{1}{72}F_{\mn}\square F^{\mn}
\biggr\rbrack \, ,
\non\\
&&\phantom{a}\non\\
{\cal L}_{\rm spin}^{(SI)} &=& -{1\over 8\pi^2} \frac{e^2}{m^2}
\biggl\lbrack
-\frac{1}{72} R F_\mn^2 
+ \frac{1}{180}R_\mn F^{\mu\alpha}F^{\nu}{}_\alpha \non\\
&&
+ \frac{1}{36}R_{\mu\nu\alpha\beta}F^\mn F^\ab
- \frac{1}{180}(\nabla_{\alpha}F_{\mn})^2
+ \frac{1}{36}F_{\mn}\square F^{\mn}
\biggr\rbrack \, .
\non\\
\label{a3SI}
\ear
Here the identities (\ref{iddFdF}) -- (\ref{idRFF}) have been used to combine some
terms. 

This is different from the heat kernel results (\ref{a2a3scal}), (\ref{a2a3spin}), which read,
after the $T$ integration,

\bear
{\cal L}_{\rm scal}^{(HK)} &=& {1\over 16\pi^2} \frac{e^2}{m^2}
\biggl\lbrack
\frac{1}{12}(\bar\xi + \frac{1}{12}) R F_\mn^2 
+ \frac{1}{90}R_\mn F^{\mu\alpha}F^{\nu}{}_\alpha \non\\
&&
- \frac{1}{60}R_{\mu\nu\alpha\beta}F^\mn F^\ab
- \frac{1}{45}(\nabla_{\alpha}F_{\mn})^2
- \frac{1}{30}F_{\mn}\square F^{\mn}\non\\
&&
-\frac{1}{180}(\nabla^{\alpha}F_{\alpha\mu})^2
\biggr\rbrack \, ,\non\\
&&\phantom{a}\non\\
{\cal L}_{\rm spin}^{(HK)} &=& -{1\over 8\pi^2} \frac{e^2}{m^2}
\biggl\lbrack
-\frac{1}{72} R F_\mn^2 
+ \frac{1}{90}R_\mn F^{\mu\alpha}F^{\nu}{}_\alpha \non\\
&&
+ \frac{1}{40}R_{\mu\nu\alpha\beta}F^\mn F^\ab
+ \frac{7}{360}(\nabla_{\alpha}F_{\mn})^2
+ \frac{1}{20}F_{\mn}\square F^{\mn}\non\\
&&
-\frac{1}{180}(\nabla^{\alpha}F_{\alpha\mu})^2
\biggr\rbrack \, .
\non\\
\label{a3DBC}
\ear
However, as expected the differences amount to total derivatives only
(see (\ref{idtd1}),(\ref{idtd2})),

\bear
{\cal L}_{\rm scal}^{(SI)} - {\cal L}_{\rm scal}^{(HK)}
&=&
{1\over 16\pi^2} \frac{e^2}{m^2}
\biggl\lbrace
\frac{7}{360}\nabla^{\alpha}(F^\mn F_{\mn;\alpha})
\non\\&&\qquad
+ \frac{1}{180}
\Bigl\lbrack
\nabla_{\alpha}(F_{\mu}{}^{\alpha}\nabla_{\beta}F^{\mu\beta})
-\nabla_{\beta}(F_{\mu}{}^{\alpha}\nabla_{\alpha}F^{\mu\beta})
\Bigr\rbrack
\biggr\rbrace \, ,
\non\\
&&\phantom{a}\non\\
{\cal L}_{\rm spin}^{(SI)} - {\cal L}_{\rm spin}^{(HK)}
&=&
-{1\over 8\pi^2} \frac{e^2}{m^2}
\biggl\lbrace
-\frac{1}{45}\nabla^{\alpha}(F^\mn F_{\mn;\alpha})
\non\\&&\qquad
+ \frac{1}{180}
\Bigl\lbrack
\nabla_{\alpha}(F_{\mu}{}^{\alpha}\nabla_{\beta}F^{\mu\beta})
-\nabla_{\beta}(F_{\mu}{}^{\alpha}\nabla_{\alpha}F^{\mu\beta})
\Bigr\rbrack
\biggr\rbrace \, .
\non\\
\ear
Similarly, agreement with the Drummond-Hathrell form of the spinor loop effective action, eq. 
(\ref{drumhath}), can be seen
using a different linear combination of the same total derivatives,

\bear
{\cal L}_{\rm spin}^{(SI)} - {\cal L}_{\rm spin}^{(DH)}
&=&
-{1\over 8\pi^2} \frac{e^2}{m^2}
\biggl\lbrace
\frac{1}{36}\nabla^{\alpha}(F^\mn F_{\mn;\alpha})
\non\\&&\qquad
+ \frac{1}{15}
\Bigl\lbrack
\nabla_{\alpha}(F_{\mu}{}^{\alpha}\nabla_{\beta}F^{\mu\beta})
-\nabla_{\beta}(F_{\mu}{}^{\alpha}\nabla_{\alpha}F^{\mu\beta})
\Bigr\rbrack
\biggr\rbrace \, .
\non\\
\ear
For completeness, let us also give here the form of the scalar loop effective
action in the Drummond-Hathrell basis:

\bear
{\cal L}_{\rm scal}^{(DH)} &=& 
{\cal L}_{\rm scal}^{(HK)}
+{1\over 16\pi^2} \frac{e^2}{m^2}
\biggl\lbrace
\frac{1}{30}\nabla^{\alpha}(F^\mn F_{\mn;\alpha})
\non\\&&\qquad\qquad\qquad
+ \frac{1}{45}
\Bigl\lbrack
\nabla_{\alpha}(F_{\mu}{}^{\alpha}\nabla_{\beta}F^{\mu\beta})
-\nabla_{\beta}(F_{\mu}{}^{\alpha}\nabla_{\alpha}F^{\mu\beta})
\Bigr\rbrack
\biggr\rbrace
\non\\
&=&
{1\over 16\pi^2} \frac{e^2}{m^2}
\biggl\lbrack
\frac{1}{12}(\bar\xi + \frac{1}{12}) R F_\mn^2 
- \frac{1}{90}R_\mn F^{\mu\alpha}F^{\nu}{}_\alpha \non\\
&&\qquad\qquad
- \frac{1}{180}R_{\mu\nu\alpha\beta}F^\mn F^\ab
+\frac{1}{60}(\nabla^{\alpha}F_{\alpha\mu})^2
\biggr\rbrack \, .
\non\\
\label{druhatscal}
\ear
The expansion (\ref{a3SI}) can be easily pursued to higher orders in $F$
using the formulas of appendix \ref{green} \cite{wip}. The reduction to
a minimal basis of terms becomes increasingly laborious, of course.

We have also checked by an independent calculation of the $a_3$ coefficients
in the DBC scheme that this scheme indeed reproduces the heat kernel results,
${\cal L}_{\rm scal,spin}^{(DBC)}={\cal L}_{\rm scal,spin}^{(HK)}$.

As was mentioned already in the introduction, Avramidi \cite{avramidi1,avramidi2}
has obtained Euler-Heisenberg type formulas for the heat kernel diagonal
of the Laplacian on twisted spin-vector bundles for the covariantly constant
case, $\nabla_{\alpha}F_{\mn} = \nabla_{\alpha}R_{\mu\nu\kappa\lambda} =  0$.
If we specialize our results (\ref{resultscal}), (\ref{resultspin}) to this case
(this just amounts to deleting all derivative terms, and in particular removes the
second integration in (\ref{resultscal}), (\ref{resultspin})) then they should
match with the result of \cite{avramidi2} after expanding to linear order in $R$ there.
However, that result is in a rather implicit form which still requires one to
perform integrals over the holonomy group for extracting individual terms in the effective action;
therefore a direct comparison would be difficult and we have not attempted it here
\footnote{Very recently Avramidi and Fucci \cite{avrfuc} have used the methods of
\cite{avramidi1,avramidi2} to obtain a more explicit representation of the heat kernel
for this covariantly constant case.}.

\section{Discussion}
\label{conclusions}
\renewcommand{\theequation}{6.\arabic{equation}}
\setcounter{equation}{0}

Let us summarize the information contained in our main result, the 
effective Lagrangians (\ref{resultscal}), (\ref{resultspin}):

\benn

\item
They contain the full information on the one -- loop amplitude involving 
$N$ photons and one graviton, with a massive scalar or spinor in the loop,
in the limit where all photon and graviton energies are small compared
to the loop mass.  In future work, we hope to obtain these amplitudes in
an explicit form, generalizing the one found for the pure $N$ -- photon 
amplitudes in \cite{mascvi}. 

\item
They can be used to extend the study of the modified dispersion relations for low-energy
photons in an Einstein-Maxwell background, previously restricted to the weak field
expansion in the electromagnetic field \cite{druhat,lapata,ditgie}, to the case of strong electromagnetic fields (although one must keep in mind that for super-strong fields 
the one-loop approximation is expected to break down already in the pure QED case 
\cite{ditgiebook}.). 

\item
It would also be straightforward to derive from  (\ref{resultscal}), (\ref{resultspin}) 
the corresponding corrections to the imaginary part of the effective
Lagrangians, in a form which modifies the Schwinger representations (\ref{ImL})
by terms of order $R/m^2$ in the prefactors of the universal exponentials.
This could be used then to calculate the pair production rates in strong electromagnetic
and weak gravitational fields. We find it hard, though, to think of a realistic scenario where
the $R/m^2$ corrections would not be negligible with respect to the leading QED term.
Here it must also be mentioned that  
Das and Dunne \cite{dasdun} have shown that the simple relation between
the imaginary part of the effective Lagrangian and the pair creation rate
does in general not extend to the curved space case.  However, this is
due to effects nonperturbative in the curvature, and is not expected to happen at
finite orders in an expansion in the curvature.

\enn

It should be emphasized that, although we have restricted ourselves here to the
approximation linear in the curvature, the formalism developed in this paper applies 
as well to the computation of the effective action at higher orders in the curvature.
The only caveat is that, starting at the quadratic level in the curvature, all of the
subtleties described in section \ref{worldline} will come into play, including the
need for a regularization and the introduction of appropriate worldline
counterterms.

Another generalization of interest would be to
consider other types of particles in the loop. Presently no natural
worldline representation is known for the case of a loop
graviton coupled to background gravity (although such a
representation can perhaps be obtained along the lines
of \cite{dahusi}). However, worldline path integrals
representing vector and antisymmetric tensor particles coupled to background
gravity have been recently constructed in \cite{babegi}.

\vskip 1cm

\noindent
{\bf Acknowledgements:}
F.B. and C.S. thank S. Theisen and the Albert-Einstein Institute, Potsdam, for hospitality
during part of this work. We also thank G.V. Dunne and U. Nucamendi for discussions, and
A. Avelino Huerta for computer help. 
The work of F.B. was supported in part by the Italian MIUR-PRIN
contract 20075ATT78. J. M. D\'avila thanks CONACYT for financial support.

\begin{appendix}

\section{Conventions and useful formulas}
\label{conventions}
\renewcommand{\theequation}{A.\arabic{equation}}
\setcounter{equation}{0}

The Einstein-Maxwell theory is described by  
\bea  
\Gamma[g,A] =    
\int d^D x\ \sqrt{g}\, \bigg (  
{1\over \kappa^2 } R - {1\over 4}F_{\mu\nu}F^{\mu\nu}  
\bigg )   
\label{EM}  
\eea  
where the metric $g_{\mu\nu}$ has signature $(-,+,+,\dots, +)$,   
$g= |{\rm det}\, g_{\mu\nu}|$, and $\kappa^2 = 16\pi G_N$.  

We use the following conventions for the curvature tensors,
\bear
[\nabla_\mu, \nabla_\nu] V^\lambda &=& 
R_{\mu\nu}{}^\lambda{}_\rho V^\rho \ , \ \ \ 
R_{\mu\nu}= R_{\lambda\mu}{}^\lambda{}_\nu 
\ , \ \  R= R^\mu{}_\mu > 0\ {\rm on\ spheres}\, , \non\\
\lbrack \nabla_\mu, \nabla_\nu \rbrack \phi &=& iF_{\mn} \phi  \, ,\non\\
\label{convcurv} 
\ear
where $V^{\mu}$ is an uncharged vector and $\phi$ a charged scalar.

The following identities are used in the text for simplifying the various effective
Lagrangians:

\bear
F_{\mu \alpha; \beta}\,F^{\mu \beta; \alpha}&=&
\frac{1}{2}F_{\mu \beta ; \alpha}\,F^{\mu \beta ; \alpha}  \, ,\label{iddFdF}\\
F^{\, \, \alpha}_{\mu}\,F^{\mu \beta}_{\, \, \ \	;\alpha \beta}&=&\frac{1}{2}F_{\mu \nu}\square F^{\mu \nu}
\, ,
\label{idFddF}\\
F_{\mu \nu}\,F_{\alpha \beta}\,R^{\mu \alpha \nu \beta}&=&
\frac{1}{2}F_{\mu \nu}\,F_{\alpha \beta}\,R^{\mu \nu \alpha \beta} \, .
\label{idRFF}
\ear
The identities (\ref{iddFdF}) -- (\ref{idRFF}) are simple consequences of
the Bianchi identities

\bear
\nabla_{\alpha} F_{\beta\gamma} 
+ \nabla_{\beta} F_{\gamma\alpha} 
+ \nabla_{\gamma} F_{\alpha\beta} &=& 0 \, ,\label{bianchiF}\\
R_{\alpha\beta\gamma\delta} 
+ R_{\beta\gamma\alpha\delta} 
+R_{\gamma\alpha\beta\delta} &=& 0 \, . \label{bianchiR}
\ear
The following identities are needed for the comparison of the various effective
Lagrangians at level $RFF$ in section 5:

\bear
\nabla^{\alpha}(F^\mn F_{\mn;\alpha})
&=& 
F_{\mn}\square F^{\mn}+ (\nabla_{\alpha} F_{\mu \nu})^{2} \, ,
\label{idtd1}\\
 \nabla_{\alpha}\left( F^{\ \alpha}_{\mu}\,\nabla_{\beta}F^{\mu \beta}\right)-\nabla_{\beta}\left( F^{\ \alpha}_{\mu}\,\nabla_{\alpha}F^{\mu \beta}\right)&=&(\nabla^{\alpha}F_{\alpha\mu})^{2} -\frac{1}{2}(\nabla_{\alpha}F_{\mu \nu})^{2}\nonumber\\
		&&+\frac{1}{2}R_{\mu\nu\alpha\beta}F^\mn F^\ab
		-R_\mn F^{\mu\alpha}F^{\nu}{}_\alpha				\, . \non\\
\label{idtd2}
\ear

\section{Properties of the field-dependent worldline Green's functions}
\label{green}
\renewcommand{\theequation}{B.\arabic{equation}}
\setcounter{equation}{0}

In this appendix we collect some basic properties of the worldline Green's functions
in a constant field ${\cal G}_B,{\cal G}_F$, introduced in (\ref{calGBGF}) 
(see appendix B of \cite{review} for a more
thorough discussion). 

${\cal G}_B$ (${\cal G}_F$) inverts the kinetic operator of the
bosonic (fermionic) parts of the 
worldline action in a background field with field strength tensor $F_\mn$. In the present conventions,
the quadratic part of the action reads

\bear
S_0[x^\mu;F] &=&  
{1\over T} \int_0^1 d\tau
\,\Bigl\lbrack\fourth\dot y^{\mu}(\tau)g_{\mn}(x_0)\dot y^{\nu}(\tau) 
+ \half iT  y^\mu (\tau) F_\mn (x_0) \dot y^\nu(\tau) \non\\
 && \qquad\qquad +\half \psi^{\mu}(\tau)g_{\mn}(x_0)\dot\psi^{\nu}(\tau) 
-iT \psi^\mu (\tau) F_\mn (x_0) \psi^\nu (\tau) 
\Bigr\rbrack \, .
\non\\
\label{S0F}
\ear
Thus formally the worldline propagators are 

\begin{eqnarray}
{\cal G}_B(\tau_1,\tau_2)&=&2
\langle \tau_1\mid
{\Bigl(
{\partial_P}^2
-2iFT\partial_P
\Bigr)}^{-1}
\mid
\tau_2\rangle \, ,\nonumber\\
{\cal G}_F(\tau_1,\tau_2)&=&2
\langle \tau_1\mid
{\Bigl(
{\partial_A}
-2iFT\Bigr)}^{-1}\mid
\tau_2\rangle \, ,
\nonumber\\
\label{calGBGFqm}
\ear
where the subscripts ``P'' and ``A'' keep track of the boundary conditions. 
Explicit formulas for these Green's functions in the
SI scheme were given already in (\ref{calGBGF}),

\bear
{\cal G}_B (\tau_1,\tau_2) &=&
{1\over 2{\cal Z}^2}\Biggl({{\cal Z}\over{{\rm sin}({\cal Z})}}
{\rm e}^{-i{\cal Z}\dot G_{B12}}
\!+\! i{\cal Z}\dot G_{B12} - 1\Biggr) \, ,
 \non\\
{\cal G}_F (\tau_1,\tau_2) &=&
G_{F12}
{{\rm e}^{-i{\cal Z}\dot G_{B12}}\over \cos ({\cal Z})} \, .
 \non
\ear
The right hand sides of these formulas are now to be understood as power series in the matrix 
${\cal Z}_\mn := TF_\mn (x_0)$, where
the indices are raised and lowered with $g_\mn (x_0)$.
%
The point of expressing ${\cal G}_{B,F}$ in terms of the ordinary worldline Green's functions
$\dot G_B, G_F$ is that it allows one to avoid making a case distinction for the ordering
of $\tau_{1,2}$.
For our present purposes also the following derivatives of ${\cal G}_{B,F}$ are needed,

\begin{eqnarray}
\dot{\cal G}_B(\tau_1,\tau_2)
&=&
{i\over {\cal Z}}\biggl({{\cal Z}\over{{\rm sin}({\cal Z})}}
\,{\rm e}^{-i{\cal Z}\dot G_{B12}}-1\biggr) \, ,
\nonumber\\
\ddot{\cal G}_{B}(\tau_1,\tau_2)
&=& 2\delta_{12} -2{{\cal Z}\over{{\rm sin}({\cal Z})}}
\,{\rm e}^{-i{\cal Z}\dot G_{B12}} \, ,\nonumber\\
\dot {\cal G}_F(\tau_1,\tau_2)
&=& 2\delta_{12} + 2iG_{F12}{{\cal Z}\over{{\rm cos}({\cal Z})}}
\, {\rm e}^{-i{\cal Z}\dot G_{B12}} \, .
\non\\
\label{derivcalGB}
\end{eqnarray}
\noindent
It will also be convenient to list the coincidence limits of the above five
functions:

\bear
{\cal G}_{B}(\tau,\tau)&=&
{1\over 2{{\cal Z}}^2}
\Bigl({\cal Z}\cot({\cal Z})-1\Bigr) \, ,\non\\
\dot {\cal G}_{B}(\tau,\tau)&=&
i\cot ({\cal Z}) - {i\over\cal Z} \, ,\non\\
\ddot {\cal G}_{B}(\tau,\tau)&=&
2\delta(0) -2{\cal Z}\cot ({\cal Z}) \, ,\non\\
{\cal G}_F(\tau,\tau) &=& -i\tan ({\cal Z})\, , \non\\
\dot{\cal G}_F(\tau,\tau) &=& 2\delta(0) + 2{\cal Z}\tan ({\cal Z}) \, .
\non\\
\label{coincidence}
\ear
\noindent
We note that ${\cal G}_B$ acts, like $G_B$, in the space of periodic
functions obeying the SI condition (\ref{defSI}). Its Fourier expansion therefore
involves only modes orthogonal to the constant functions. This has the consequence
that 

\bear
\int_0^1d\tau_{1,2} \,{\cal G}_B^{(n)}(\tau_1,\tau_2) &= & 0 \, ,
\label{intort}
\ear
where $n$ denotes any derivative of ${\cal G}_B$. 
Finally, to recover perturbative results one needs 
the coefficients in the expansions of 
${\cal G}_{B,F}$ as powers of $F$. These expansions can be written as follows,

\bear
{\cal G}_B(\tau_1,\tau_2) &=& -2\sum_{n=0}^{\infty} (2i{\cal Z})^n g_{n+2}(\tau_1-\tau_2) \, ,
\non\\
\dot{\cal G}_B(\tau_1,\tau_2) &=& -2\sum_{n=0}^{\infty} (2i{\cal Z})^n g_{n+1}(\tau_1-\tau_2) \, ,
\non\\
\ddot{\cal G}_B(\tau_1,\tau_2) &=& 2\delta_{12} - 2\sum_{n=0}^{\infty} (2i{\cal Z})^n
g_n(\tau_1-\tau_2) \, ,
\non\\
{\cal G}_F(\tau_1,\tau_2) &=& 2\sum_{n=0}^{\infty} (2i{\cal Z})^n f_{n+1}(\tau_1-\tau_2) \, ,
\non\\
\dot{\cal G}_F(\tau_1,\tau_2) &=& 2\delta_{12} + 2\sum_{n=1}^{\infty} (2i{\cal Z})^n
f_n(\tau_1-\tau_2) \, .
\non\\
\label{expGBGFgeneral}
\ear
Here the coefficient functions $g_n,f_n$ are polynomials in $\tau_1-\tau_2$
(apart from factors of ${\rm sign}(\tau_1-\tau_2)$). In writing these polynomials 
one has a choice of variables. In terms of $\tau=\tau_1-\tau_2$ one gets, by a straightforward
expansion of (\ref{calGBGF}) (see \cite{ss3,review}),

\bear
g_n(\tau) &=& {1\over n!} {\cal B}_n(\abs{\tau}){\rm sign}^n(\tau) \, ,\non\\
f_n(\tau) &=& {1\over 2(n-1)!} {\cal E}_{n-1}(\abs{\tau}){\rm sign}^n(\tau) \, .\non\\
\label{gfbernoulli}
\ear
Here ${\cal B}_n$ denotes the $n$th Bernoulli polynomial, ${\cal E}_n$ the $n$th
Euler polynomial.

Alternatively, one can also 
write the same coefficient functions in terms of the vacuum Green's
functions \cite{dhrs}. 
Denoting
by $\bar G$ the coordinate worldline Green's function
with its coincidence limit subtracted,

\bear
\bar G(\tau) &:=& \abs{\tau} - \tau^2 
\label{defGbar}
\ear
one finds 

\bear
g_0(\tau) &=& 1 \, ,\non\\
g_1(\tau) &=& -\half \dot G_B(\tau,0)=-\half \dot{\bar G}(\tau) \, ,\non\\
g_2(\tau) &=& -\half G_B(\tau,0) = -\half \bar G(\tau) +{1\over 12} \, ,\non\\
g_n(\tau) &=& 
\left\{\begin{array}{cc}
     {B_n\over n!} + {1\over 2(n-1)!}
     \sum_{k=1}^{n/2-1}f\bigl({n\over 2}-1,k\bigr)(-\bar G)^{k+1}(\tau) & (n>2 \quad {\rm even}) \\
    -{1\over 2 n!}  \sum_{k=1}^{(n-1)/2}
    f\bigl({n-1\over 2},k\bigr)(k+1) \dot{\bar G}(\tau) (-\bar G)^k(\tau)
 & (n>2 \quad {\rm odd})\\
\end{array}
\right.\non\\
\label{gfaulhaber}
\ear
and

\bear
f_1(\tau) &=&\half G_F(\tau,0) = \half {\rm sign}(\tau) \, ,\non\\
f_2(\tau) &=&  -\fourth G_F(\tau,0)\dot G_B(\tau,0)=-\fourth {\rm sign}(\tau)\dot{\bar G}(\tau) \, ,\non\\
f_n(\tau) &=& 
\left\{\begin{array}{cc}
   - {1\over 2 n!}
     \sum_{k=1}^{n/2}s\bigl({n\over 2},k\bigr)k\,
     {\rm sign}(\tau)\dot{\bar G}(\tau)(-\bar G)^{k-1}(\tau) 
     & (n>2 \quad {\rm even}) \\
{1\over 2(n-1)!}\sum_{k=1}^{(n-1)/2}s
\bigl({n-1\over 2},k\bigr){\rm sign}(\tau)(-\bar G)^k(\tau) & (n>2 \quad {\rm odd}) \, .\\
\end{array}
\right.\non\\
\label{fsalie}
\ear
Here the $f(m,k)$ are Faulhaber numbers and the $s(m,k)$  
Sali\'e numbers. Those numbers can be defined in terms of the
Bernoulli numbers as 
\cite{gesvie}

\bear
f(m,k) &=& (-1)^{k+1}\sum_{j=0}^{\lfloor(k-1)/2\rfloor}
{1\over k-j}{2k-2j\choose k+1}{2m+1\choose 2j+1}B_{2m-2j} \, ,
\non\\
s(m,k) &=& 2(-1)^k\sum_{j=0}^{\lfloor(k-1)/2\rfloor}
{1\over 2k-2j-1}{2k-2j-1\choose k}{2m\choose 2j}\Bigl(1-2^{2m-2j}\Bigr)B_{2m-2j}
\non\\
\label{deffaulsal}
\ear
($m\geq k \geq 1$).
For easy reference, let us write down the expansions
(\ref{expGBGFgeneral}) explicitly to order $O(F^2)$, using the form (\ref{gfaulhaber}), (\ref{fsalie})
for the coefficients:

\begin{eqnarray}
{\cal G}_{B12} &=& {\bar G}_{B12}-{1\over 6}
-{i\over 3}
\dot G_{B12}{\bar G}_{B12}{\cal Z} + ({1\over 3}{\bar G}_{B12}^2
-{1\over 90}){\cal Z}^2+O({\cal Z}^3) \, ,\nonumber\\
\dot{\cal G}_{B12} 
&=&\dot G_{B12}+2i\Bigl({\bar G}_{B12}-{1\over 6}\Bigr){\cal Z}
+{2\over 3}\dot G_{B12}{\bar G}_{B12}{\cal Z}^2 +  O({\cal Z}^3) \, ,
\nonumber\\
\ddot{\cal G}_{B12} 
&=& 2 \delta_{12} - 2 +2i\dot G_{B12}{\cal Z}
-4\Bigl({\bar G}_{B12}-{1\over 6}\Bigr){\cal Z}^2+O({\cal Z}^3) \, ,\nonumber\\
{\cal G}_{F12}&=& G_{F12}-iG_{F12}\dot G_{B12}{\cal Z}
+2G_{F12}{\bar G}_{B12}{\cal Z}^2+O({\cal Z}^3) \, ,\nonumber\\
\dot{\cal G}_{F12}&=& 2\delta_{12} +2i{\cal Z}G_{F12}
+2G_{F12}\dot G_{B12}{\cal Z}^2 + O({\cal Z}^3) \, .\nonumber\\
\label{GB(F)expand}
\end{eqnarray}

\end{appendix}

%
%


\end{document}